\documentclass[10pt,journal]{IEEEtran}

\pdfoutput=1
\usepackage{graphicx}
\usepackage{amsmath,amsxtra, amsfonts, amssymb}
\usepackage{setspace}
\usepackage{times}
\usepackage{epsfig}
\usepackage{latexsym}
\usepackage{cite}
\usepackage{url}
\usepackage{epstopdf}
\usepackage{tikz,pgf,pgfplots}
\usepackage{multirow}
\usepackage{acronym}
\usepackage{mathtools} 
\usepackage{subcaption}
\usepackage{flushend}

\newtheorem{remark}{Remark}

\newcommand{\pv}{\boldsymbol{p}}

\newcommand{\SN}{{\rm S}} 
\newcommand{\RN}{{\rm R}} 
\newcommand{\DN}{{\rm D}} 
\newcommand{\iN}{i} 
\newcommand{\jN}{j} 
\newcommand{\PN}{{\rm P}} 
\newcommand{\RSet}{{\mathcal{R}}} 
\newcommand{\KSet}{{\mathcal{K}}} 

\newcommand{\hsd}{h_{\SN \DN}}
\newcommand{\hij}{h_{ij}}
\newcommand{\hsk}{h_{\SN k}}
\newcommand{\hkk}{h_{kk}}
\newcommand{\hkd}{h_{k \DN}}

\newcommand{\gsd}{g_{\SN \DN}}
\newcommand{\gij}{g_{ij}}
\newcommand{\gsk}{g_{\SN k}}
\newcommand{\gkk}{g_{kk}}
\newcommand{\gkd}{g_{k \DN}}
\newcommand{\gsp}{g_{\SN \PN}}
\newcommand{\gkp}{g_{k \PN}}

\newcommand{\pisr}{{\pi_{\SN \RN}}}
\newcommand{\pirr}{{\pi_{\RN \RN}}}
\newcommand{\pird}{{\pi_{\RN \DN}}}
\newcommand{\pisd}{{\pi_{\SN \DN}}}

\newcommand{\piij}{{\pi_{ij}}}

\newcommand{\pisp}{{\pi_{\SN \PN}}}

\newcommand{\pirp}{{\pi_{\RN \PN}}}

\newcommand{\msr}{{m_{\SN \RN}}}
\newcommand{\mrr}{{m_{\RN \RN}}}
\newcommand{\mrd}{{m_{\RN \DN}}}
\newcommand{\msd}{{m_{\SN \DN}}}

\newcommand{\mij}{{m_{ij}}}
\newcommand{\msk}{{m_{\SN k}}}
\newcommand{\mkk}{{m_{kk}}}
\newcommand{\mkd}{{m_{k \DN}}}
\newcommand{\msp}{{m_{\SN \PN}}}
\newcommand{\mkp}{{m_{k \PN}}}
\newcommand{\mrp}{{m_{\RN \PN}}}

\newcommand{\thetasr}{{\theta_{\SN \RN}}}
\newcommand{\thetarr}{{\theta_{\RN \RN}}}
\newcommand{\thetard}{{\theta_{\RN \DN}}}
\newcommand{\thetasd}{{\theta_{\SN \DN}}}

\newcommand{\thetaij}{{\theta_{ij}}}
\newcommand{\thetask}{{\theta_{\SN k}}}
\newcommand{\thetakk}{{\theta_{kk}}}
\newcommand{\thetakd}{{\theta_{k \DN}}}
\newcommand{\thetasp}{{\theta_{\SN \PN}}}
\newcommand{\thetakp}{{\theta_{k \PN}}}
\newcommand{\thetarp}{{\theta_{\RN \PN}}}

\newcommand{\GAMMA}{\mathcal{G}}
\newcommand{\RV}{X}
\newcommand{\RVval}{x}

\DeclarePairedDelimiterX\GAMFN[1]{}{}{\boldsymbol{\Gamma}\left(#1\right)}
\DeclarePairedDelimiterX\UPRGAMFN[1]{}{}{\boldsymbol{\Gamma}\left(#1\right)}
\DeclarePairedDelimiterX\LWRGAMFN[1]{}{}{\boldsymbol{\gamma}\left(#1\right)}
\DeclarePairedDelimiterX\WTKRFN[3]{}{}{{\boldsymbol{W}}_{#1,#2}\left( #3 \right)}
\DeclarePairedDelimiterX\KMRFN[3]{}{}{\boldsymbol{\left._1F_1\right.} \left( #1;#2;#3 \right)}
\DeclarePairedDelimiterX\BTAFN[2]{}{}{\boldsymbol{\mathcal{B}} \left( #1,#2 \right)}
\DeclarePairedDelimiterX\PDF[2]{}{}{\boldsymbol{f}_{#1} \left( #2 \right)}
\DeclarePairedDelimiterX\CDF[2]{}{}{\boldsymbol{F}_{#1} \left( #2 \right)}
\DeclarePairedDelimiterX\TRICOMIFN[3]{}{}{\boldsymbol{U}\left(#1, #2, #3 \right)}

\newcommand{\XS}{x_{\SN}}

\newcommand{\Xk}{x_{k}}

\newcommand{\Yk}{y_{k}}
\newcommand{\YD}{y_{\DN}}

\newcommand{\Nk}{n_k}
\newcommand{\ND}{n_{\DN}}
\newcommand{\XI}{x_i}
\newcommand{\NI}{n_i}

\newcommand{\PS}{P_{\SN}}						
\newcommand{\PR}{P_{\RN}}						
\newcommand{\PI}{P_{i}}
\newcommand{\PWRSCL}{\delta}					
\newcommand{\Pgen}{P}							
\newcommand{\Pk}{P_{k}}						

\newcommand{\SkFDSNR}{\gamma_{\SN k}}
\newcommand{\SkDFDSNR}{\gamma_{(\SN, k)\rightarrow \DN}}
\newcommand{\kDFDSNR}{\gamma_{k \DN}}
\newcommand{\kFDSNR}{\gamma_{k}}
\newcommand{\RDFDSNR}{\gamma_{\RN \DN}}
\newcommand{\SDFDSNR}{\gamma_{\SN \DN}}

\newcommand{\etoeFDSNR}{\gamma_{\rm e2e}}

\newcommand{\Rate}{R}  
\newcommand{\SNRFD}{\eta}  
\newcommand{\TM}{t}	

\newcommand{\Pout}{{\mathcal{P}_{\rm out}}}

\newcommand{\Ith}{I_{\rm th}}
\newcommand{\Isp}{I_{\SN \PN}}
\newcommand{\Ikp}{I_{k \PN}}
\newcommand{\Isk}{I_{k}}

\newcommand{\PL}{\mathcal{P}_L}
\newcommand{\Pzero}{\mathcal{P}_0}

\newcommand{\CDFNDL}{F_{\etoeFDSNR}^{\text{NDL}}(\RVval)}
\newcommand{\PNDL}{\mathcal{P}^{\text{NDL}}_{\text{out}}}
\newcommand{\DNDL}{\mathcal{D}^{\text{NDL}}}
\newcommand{\CDFIDL}{F_{\etoeFDSNR}^{\text{IDL}}(\RVval)}
\newcommand{\PIDL}{\mathcal{P}^{\text{IDL}}_{\text{out}}}
\newcommand{\DIDL}{\mathcal{D}^{\text{IDL}}}
\newcommand{\CDFMHDT}{F_{\etoeFDSNR}^{\text{IDL/DT}}(\RVval)}
\newcommand{\PMHDT}{\mathcal{P}^{\text{IDL/DT}}_{\text{out}}}
\newcommand{\DMHDT}{\mathcal{D}^{\text{IDL/DT}}}
\newcommand{\PSDFFDRS}{\mathcal{P}^{\text{SDF}}_{\text{out}}}
\newcommand{\DSDFFDRS}{\mathcal{D}^{\text{SDF}}}

\newcommand{\TPT}{\mathcal{T}}

\newcommand{\nvarD}{\sigma_{\DN}^2}
\newcommand{\nvark}{\sigma_{k}^2}

\acrodef{5G}{fifth-generation}
\acrodef{ACK}{acknowledgement}
\acrodef{AWGN}{additive white {G}aussian noise}
\acrodef{AF}{amplify-and-forward}
\acrodef{bpcu}{bits per channel use}
\acrodef{CDF}{cumulative distribution function}
\acrodef{CEMSE}{Computer, Electrical and Mathematical Science and Engineering Division}
\acrodef{CRN}{Cognitive relay network}
\acrodef{CSI}{channel state information}
\acrodef{DF}{decode-and-forward}
\acrodef{DT}{direct transmission}
\acrodef{FDR}{full-duplex relaying}
\acrodef{FDRS}{full-duplex relay selection}
\acrodef{HDR}{half-duplex relaying}
\acrodef{HDRS}{half-duplex relay selection}
\acrodef{IDF}{incremental decode-and-forward}
\acrodef{IDL}{interfering direct link}
\acrodef{iid}[i.i.d.]{independent and identically distributed}
\acrodef{inid}[i.n.i.d.]{independent but not identically distributed}
\acrodef{KAUST}{King Abdullah University of Science and Technology}
\acrodef{LoS}{line-of-sight}
\acrodef{MHDF}{multi-hop decode-and-forward}
\acrodef{MISO}{multiple-input single-output}
\acrodef{MRC}{maximum-ratio combining}
\acrodef{NDL}{no direct link}
\acrodef{PDF}{probability density function}
\acrodef{RV}{random variable}
\acrodef{RSI}{residual self-interference}
\acrodef{SDF}{selective DF}
\acrodef{SINR}{signal-to-interference-plus-noise ratio}
\acrodef{SNR}{signal-to-noise ratio}

\newcommand{\figscl}{0.7}
\newcommand{\figscal}{0.8}

\usepackage{etoolbox}
\allowdisplaybreaks

\begin{document}
\title{Full-Duplex Relay Selection \\in Cognitive Underlay Networks}
\author{Mohammad~Galal~Khafagy,~\IEEEmembership{Senior Member,~IEEE,}
        Mohamed-Slim~Alouini,~\IEEEmembership{Fellow,~IEEE,}
        \\and~Sonia~A\"issa, ~\IEEEmembership{Senior Member,~IEEE} \vspace{-4mm}
\thanks{The research reported in this publication was supported by funding from \ac{KAUST}.} 
\thanks{This work was done when M. G. Khafagy was with \ac{KAUST}, \ac{CEMSE}, Thuwal, Makkah Province, Saudi Arabia (e-mail: mohammad.khafagy@kaust.edu.sa). He is currently affiliated with the Computer Science Department, College of Engineering, Qatar University, Doha, Qatar. M.-S. Alouini is with \ac{KAUST}, \ac{CEMSE}, Thuwal, Makkah Province, Saudi Arabia (e-mail: slim.alouini@kaust.edu.sa). S. A\"issa is with the Institut National de la Recherche Scientifique (INRS), University of Quebec, Montreal, QC, H5A 1K6, Canada (e-mail: aissa@emt.inrs.ca).}
}
\maketitle

\makeatletter
\newif\ifAC@uppercase@first%
\def\Aclp#1{\AC@uppercase@firsttrue\aclp{#1}\AC@uppercase@firstfalse}%
\def\AC@aclp#1{%
  \ifcsname fn@#1@PL\endcsname%
    \ifAC@uppercase@first%
      \expandafter\expandafter\expandafter\MakeUppercase\csname fn@#1@PL\endcsname%
    \else%
      \csname fn@#1@PL\endcsname%
    \fi%
  \else%
    \AC@acl{#1}s%
  \fi%
}%
\def\Acp#1{\AC@uppercase@firsttrue\acp{#1}\AC@uppercase@firstfalse}%
\def\AC@acp#1{%
  \ifcsname fn@#1@PL\endcsname%
    \ifAC@uppercase@first%
      \expandafter\expandafter\expandafter\MakeUppercase\csname fn@#1@PL\endcsname%
    \else%
      \csname fn@#1@PL\endcsname%
    \fi%
  \else%
    \AC@ac{#1}s%
  \fi%
}%
\def\Acfp#1{\AC@uppercase@firsttrue\acfp{#1}\AC@uppercase@firstfalse}%
\def\AC@acfp#1{%
  \ifcsname fn@#1@PL\endcsname%
    \ifAC@uppercase@first%
      \expandafter\expandafter\expandafter\MakeUppercase\csname fn@#1@PL\endcsname%
    \else%
      \csname fn@#1@PL\endcsname%
    \fi%
  \else%
    \AC@acf{#1}s%
  \fi%
}%
\def\Acsp#1{\AC@uppercase@firsttrue\acsp{#1}\AC@uppercase@firstfalse}%
\def\AC@acsp#1{%
  \ifcsname fn@#1@PL\endcsname%
    \ifAC@uppercase@first%
      \expandafter\expandafter\expandafter\MakeUppercase\csname fn@#1@PL\endcsname%
    \else%
      \csname fn@#1@PL\endcsname%
    \fi%
  \else%
    \AC@acs{#1}s%
  \fi%
}%
\edef\AC@uppercase@write{\string\ifAC@uppercase@first\string\expandafter\string\MakeUppercase\string\fi\space}%
\def\AC@acrodef#1[#2]#3{%
  \@bsphack%
  \protected@write\@auxout{}{%
    \string\newacro{#1}[#2]{\AC@uppercase@write #3}%
  }\@esphack%
}%
\def\Acl#1{\AC@uppercase@firsttrue\acl{#1}\AC@uppercase@firstfalse}
\def\Acf#1{\AC@uppercase@firsttrue\acf{#1}\AC@uppercase@firstfalse}
\def\Ac#1{\AC@uppercase@firsttrue\ac{#1}\AC@uppercase@firstfalse}
\def\Acs#1{\AC@uppercase@firsttrue\acs{#1}\AC@uppercase@firstfalse}
\robustify\Aclp
\robustify\Acfp
\robustify\Acp
\robustify\Acsp
\robustify\Acl
\robustify\Acf
\robustify\Ac
\robustify\Acs
\makeatother

\begin{abstract}
In this work, we analyze the performance of \ac{FDRS} in spectrum-sharing networks. Contrary to half-duplex relaying, \ac{FDR} enables simultaneous listening/forwarding at the secondary relay(s), thereby allowing for a higher spectral efficiency. However, since the source and relay simultaneously transmit in \ac{FDR}, their superimposed signal at the primary receiver should now satisfy the existing interference constraint, which can considerably limit the secondary network throughput. In this regard, relay selection can offer an adequate solution to boost the secondary throughput while satisfying the imposed interference limit. We first analyze the performance of opportunistic \ac{FDRS} with \ac{RSI} by deriving the exact cumulative distribution function of its end-to-end signal-to-interference-plus-noise ratio under Nakagami-$m$ fading. We also evaluate the offered diversity gain of relay selection for different full-duplex cooperation schemes in the presence/absence of a direct source-destination link.  When the adopted \ac{RSI} link gain model is sublinear in the relay power, which agrees with recent research findings, we show that remarkable diversity gain can be recovered even in the presence of an interfering direct link. Second, we evaluate the end-to-end performance of \ac{FDRS} with interference constraints due to the presence of a primary receiver. Finally, the presented exact theoretical findings are verified by numerical simulations.
\end{abstract}
\begin{keywords}
diversity gain, full-duplex, opportunistic relay selection, outage performance, self-interference, spectrum sharing.
\end{keywords}

\acresetall

\section{Introduction}
\IEEEPARstart{S}{pectrum} sharing represents one paradigm for the celebrated cognitive radio technology in which adequate means are offered to resolve the scarcity problem of wireless resources \cite{199908_PC_Cognitive_Mitola,200905_Proc_Cognitive_Goldsmith}. Also known as underlay cognitive radio, spectrum sharing  allows the traffic of secondary \emph{cognitive} users to coexist with that of the primary spectrum users as long as a certain interference level is not exceeded. In particular, when the secondary source and destination are spatially isolated, relay assistance becomes inevitable to establish successful communication via relay listening/forwarding, while satisfying the coexistence constraints with the primary user. 

\acp{CRN} continue to draw a noticeable interest in the wireless communications community for their coverage extension capabilities under an efficient spectrum usage \cite{200905_ProcIEEE_CRNs}. Adhering to the interference constraints in underlay networks, however, can considerably limit the throughput of the secondary system, especially when a single relay is leveraged to assist the communication between the secondary source and its far destination. To tackle such a challenge, relay selection
\cite{200603_JSAC_Bletsas_Relay_selection,200810_TWC_Michlopoulos_Relay_Selection}
was introduced to underlay \acp{CRN}, and it was shown to offer remarkable performance gains relative to its fixed relaying counterpart. Cognitive relay selection was fairly investigated in the literature under the relaying strategies of \ac{AF} \cite{201206_ICC_Hussain_AF1,201308_WCL_Chen_AF}
and \ac{DF} \cite{201102_TWC_Lee_DF,201110_TWC_Sagong_DF}.

All the aforementioned efforts considered the conventional \ac{HDR}. However, \ac{HDR} suffers from a spectral efficiency loss when compared to \ac{DT} due to its time-orthogonal relay listening/forwarding. More recently, full-duplex communications experienced serious developments that brought the first full-duplex node prototypes into reality \cite{mobicom2011fullduplex,201212_TWC_Duarte}.
These developments motivated further research in cooperative contexts where simultaneous listening/forwarding was allowed by \emph{\ac{FDR}}, thereby eliminating the known rate loss of \ac{HDR}. Since then, recent efforts were directed to study FDR in several directions, for instance, in \acp{CRN} \cite{201305_TWC_Cognitive_FDR_Krikidis,201210_TWC_Cognitive_FDR_Daesik_DF},
and with relay selection \cite{201206_ICC_Krikidis_FDR_Selection_AF,201212_TWC_Krikidis_FDR_Selection_AF,201012_EL_Rui_FDR_Selection_DF,201406_ICC14_FDRS_Zhong,Khafagy_CogFDRS_ICCW15, Khafagy_FDRS_CAMAD15, 201505_Hanzo_RS_Cognitive, 201510_TVT_FDRS_Zhong, 201602_IET_Wang_FDRS, 201701_SigTelCom, 2017XX_SurvTut_RS_Cognitive}. These research endeavors and practical potentials have motivated the International Telecommunication Union (ITU) to recommend FDR as one of two full-duplex communication settings for employment in future \ac{5G} cellular networks \cite{201509_ITU_R_RECOMMENDATION_M2083,201411_ITU_R_REPORT_M2320}.

\subsection{Related Work}
In \cite{201210_TWC_Cognitive_FDR_Daesik_DF}, a cognitive underlay setting is studied in which a secondary network shares the spectrum with a primary network under a certain interference constraint. The secondary system comprises a source, a destination and a full-duplex DF relay. Under the aforementioned interference constraint, optimal source and relay power allocation was investigated with the objective to minimize the end-to-end outage probability. In \cite{201206_ICC_Krikidis_FDR_Selection_AF,201212_TWC_Krikidis_FDR_Selection_AF}, Krikidis \emph{et al.} studied the outage performance of \ac{AF} \ac{FDRS} with \ac{RSI}, when no direct link exists between the source and the destination. For \ac{DF} networks, \ac{FDRS} was briefly discussed in \cite{201012_EL_Rui_FDR_Selection_DF}, however only the simple case of non-fading \ac{RSI} link was considered, and the adverse effect of the interfering direct link in \ac{FDR} was not accounted for. In \cite{201406_ICC14_FDRS_Zhong}, \ac{DF} \ac{FDRS} was studied under Rayleigh fading while accounting for the effect of a fading \ac{RSI} link, yet the direct link was assumed unavailable. In \cite{Khafagy_CogFDRS_ICCW15}, the authors have studied \ac{DF} \ac{FDRS} under Rayleigh fading while jointly taking into account the effects of fading \ac{RSI} and direct links, in an underlay setting where a primary receiver imposes an interference constraint. The authors also analyzed the performance of \ac{DF} \ac{FDRS} in \cite{Khafagy_FDRS_CAMAD15} under Nakagami-$m$ fading in the absence of spectrum sharing. A similar setting was also briefly  discussed in \cite{201505_Hanzo_RS_Cognitive} for Nakagami-$m$ fading, where the relay selection is performed based only on the strength of the self-interference channel. However, the direct source-destination link was assumed unavailable  in \cite{201505_Hanzo_RS_Cognitive}, and also no analysis was provided. The results in \cite{201406_ICC14_FDRS_Zhong} have been recently extended in \cite{201510_TVT_FDRS_Zhong} to Nakagami-$m$ fading scenarios when no direct link exists in the presence of a primary receiver. Also recently, \cite{201602_IET_Wang_FDRS} has studied the extension of the \ac{SDF} \ac{FDR} protocol proposed by the authors in \cite{6510556,2015XX_TWC_Khafagy_FDR} to the \ac{FDRS} setting under Nakagami-$m$ fading without spectrum sharing constraints. The work in\cite{201701_SigTelCom} considered partial relay selection, based on the link strengths of only the first hop, under Rayleigh fading and without a direct source-destination link when multiple primary users exist. The interested reader is also referred to \cite{2017XX_SurvTut_RS_Cognitive} for a recent comprehensive survey, listing the latest research endeavors on the topic.

\begin{figure}[!t]
\centering
\includegraphics[width=\figscal\columnwidth]{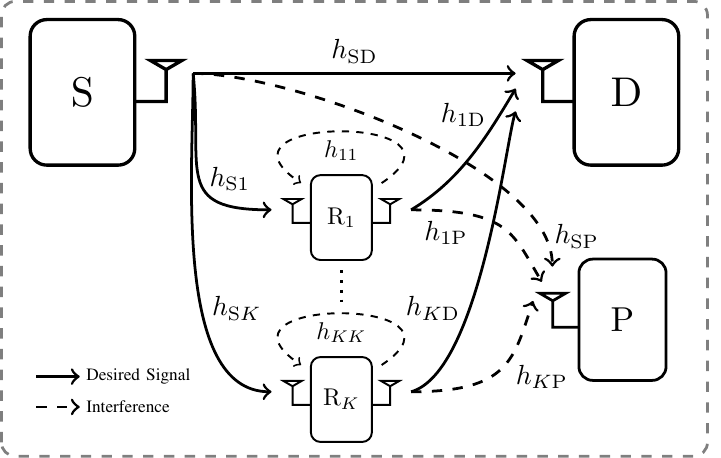}
\caption{A spectrum-sharing network with full-duplex relaying.}
\label{sysmodfig_CogFDRS}
\vspace{-3mm}
\end{figure}

\subsection{Contribution}
Despite its anticipated performance merits, the incorporation of relay selection techniques into cognitive \ac{FDR} networks remained untackled to the best of the authors' knowledge till the work by the authors in \cite{Khafagy_CogFDRS_ICCW15}. In this work, we aim at bridging this gap by introducing relay selection to cognitive \ac{FDR} networks and analyzing its offered performance gains. In underlay settings with \ac{FDR}, the performance of the secondary users can be seriously limited due to the fact that now the superimposed source and relay interference components should satisfy the existing interference constraint which was previously imposed in \ac{HDR} on each transmit node at a time. In this regard, relay selection techniques can offer an adequate solution to boost the secondary network throughput. Moreover, in full-duplex systems, the analysis of the end-to-end diversity gain highly depends on the proper modeling of the \ac{RSI} link. Based on the available cancellation techniques at that time, earlier modeling considered the RSI link gain to scale linearly with the transmit power of the full-duplex node \cite{201307_TWC_Elsayed_FD}. More recently, new passive suppression and active analog/digital cancellation techniques were suggested and analyzed \cite{201308_SIGCOMM_FD_Radios,2015XX_TWC_Elsayed_FD1,2015XX_TWC_Elsayed_FD2}. These techniques propose a different scaling trend, where the adaptive cancellation methods may actually succeed to yield a nearly constant level of the average RSI link gain while varying the relay power. Thus, considering a model that spans the whole range of linear, sublinear and constant power scaling is crucial for the performance evaluation of \ac{FDR}. This general model has been adopted in the recent literature \cite{6832455,2015XX_TWC_Khafagy_FDR}. 

Our contribution in this paper is summarized as follows:\footnote{While this work was in progress, preliminary results were accepted for presentation in IEEE ICC Wkshps'15 \cite{Khafagy_CogFDRS_ICCW15} for the spectrum-sharing setting under Rayleigh fading, and in IEEE CAMAD'15 \cite{Khafagy_FDRS_CAMAD15} for the non-spectrum-sharing setting under Nakagami-$m$ fading. The currently presented work reports the full results for the spectrum-sharing setting under Nakagami-$m$ fading which subsumes the results in \cite{Khafagy_CogFDRS_ICCW15} and \cite{Khafagy_FDRS_CAMAD15} as special cases. Also presented is the diversity analysis of the \ac{FDRS} protocols, as well as the relative performance to the \ac{HDRS} protocols. }

\begin{itemize}
\item We derive the exact \ac{CDF} for the end-to-end \ac{SINR} of opportunistic \ac{FDRS} in the presence/absence of a direct source-destination link in addition to a spectrum-sharing constraint, where all links experience Nakagami-$m$ fading. The end-to-end \ac{SINR} \ac{CDF} is obtained and compared for two cooperation classes, namely, 1) multihop cooperation \cite{TVT2010_DF_FDR_RSI_without_DL}, where information is distilled only from the relayed path, and 2) selective block cooperation \cite{6510556,2015XX_TWC_Khafagy_FDR,201602_IET_Wang_FDRS}, where the direct information from the source is additionally leveraged at the destination. When a direct source-destination link exists, the \ac{SINR} values of the different relaying paths become mutually dependent, which we account for in the analysis. Also, \ac{DT} can take place if the direct link supports higher instantaneous rate than those offered through relaying.
\item From the derived \ac{CDF} expressions, we analyze the diversity gains offered by \ac{FDRS} in the presence/absence of direct source-destination link. When the adopted \ac{RSI} link gain model is sublinear in the relay power, which agrees with recent research findings \cite{201308_SIGCOMM_FD_Radios,2015XX_TWC_Elsayed_FD1,2015XX_TWC_Elsayed_FD2}, we show that a remarkable diversity gain can be achieved even for the simpler multihop cooperation protocols which treat the direct link as interference to the relayed paths. This result gets around a known drawback in FDR in the recent literature where a zero diversity order, i.e., an error floor, was encountered even in the absence of \ac{RSI} due to the interfering direct link adverse effect. We also show that hybrid multihop-based \ac{FDRS}/\ac{DT} achieves the same diversity order of the \ac{SDF}-\ac{FDRS}. Since multihop cooperation adopts a simpler end-to-end transmission/decoding approach, it may be favored over the class of selective cooperation when decoding complexity represents a major concern.
\item We compare the performance of the considered full-duplex cooperative schemes to that of their half-duplex counterparts for different scaling trends of the \ac{RSI}, also in the presence of a primary receiver.
\end{itemize}
\section{System Model}
We consider the underlay cognitive setting depicted in Fig. \ref{sysmodfig_CogFDRS}. As shown, a secondary source ($\SN$) intends to communicate with a secondary destination ($\DN$) in the presence of a primary receiver ($\PN$). In agreement with similar recent studies of underlay cognitive settings, for instance in \cite{201102_TWC_Lee_DF}, the primary source is assumed to be far enough that its interference effect on the secondary system can be reasonably ignored. The direct secondary source-destination link is assumed of a relatively weak gain due to distance and shadowing effects. Hence, a full-duplex relay is utilized to assist the end-to-end secondary communication, taking the interference constraint on the primary receiver into account. Although \ac{FDR} can offer higher spectral efficiency when compared to its half-duplex counterpart, it introduces an additional challenge in cognitive settings where higher interference levels may be experienced by the primary user of the spectrum band, due to simultaneous source/relay transmissions. Also, \ac{FDR} suffers from an \ac{RSI} level which imposes an additional communication challenge.

In this regard, relay selection is proposed as an adequate mean to alleviate the aforementioned interference problems. Specifically, the secondary system has the additional degree of freedom to select one relay out of a cluster of $K$ full-duplex relays, $\RSet=\{\RN_1, \RN_2, \cdots, \RN_K\}$. Relay selection is widely known to offer the same diversity order of scenarios where the transmissions of multiple cooperating relays are combined, yet it has the advantage of maintaining a simple transmitter/receiver structure. In cognitive scenarios, relay selection remains of a particular payoff as it offers the ability to pick the best relay in terms of a certain performance criterion while maintaining the interference limit on the primary user.

The relay selection process can be governed either via a distributed \cite{200603_JSAC_Bletsas_Relay_selection} or a centralized fashion \cite{200810_TWC_Michlopoulos_Relay_Selection}. For example, a distributed scheme can be possibly adopted for the considered setting, where each relay estimates the receive \ac{CSI} of its own incoming link from the source. Also, it estimates the \ac{CSI} of its outgoing link from the previously received \ac{ACK} packet from the destination while leveraging channel reciprocity. Accordingly, a local reverse timer is triggered at each relay based on its end-to-end channel quality, where the better the channel condition is, the faster the timer expires. All relays operate in a listening mode within this phase, allowing the relay with the first expired timer to transmit, while all other relays accordingly backoff. Such a reverse timer/backoff method is widely known for distributed wireless multiple access purposes \cite{200603_JSAC_Bletsas_Relay_selection}. When a direct source-destination link exists, the destination is assumed to additionally feedback its \ac{CSI} within the \ac{ACK} packet of the previous transmission. Since an underlay setting is considered, no cooperation is assumed between the primary and secondary networks. Nonetheless, the relays are assumed to overhear the \ac{ACK} packets sent from the primary receiver to its own source, and hence, locally estimate its channel, also utilizing channel reciprocity. It can be easily noticed that only local link strengths are required here, aided by simple feedback from the destination. Hence, the overhead does not scale with the number of relays. It should be also noted here that this knowledge is only required for the relay selection at the upper layers of the communication protocol stack. Such a knowledge is not required for the communication process itself over the physical layer. Hence,  further enhancements to the relay selection process at the upper layers remain possible, however, this lies beyond the scope of the current paper.
%
\subsection{Channel Model}\label{subsec:chan_model}
The fading coefficient of the $\iN - \jN$ link is denoted by $\hij$, for $\iN \in\{\SN,\KSet\}$ and $\jN \in\{\KSet,\DN,\PN\}$, where $\KSet=\{1,2,\cdots,K\}$ denotes the set of relay indices. The $\iN - \jN$ link gain is denoted by $\gij= \vert \hij \vert ^2$. The relays operate in full-duplex mode where simultaneous listening/forwarding is allowed with an introduced level of loopback interference. The link gain $\hkk, \forall k \in \KSet$, represents the \ac{RSI} after undergoing all possible isolation and cancellation techniques, as for instance in \cite{riihonen201109TWC}. All channels are assumed to follow a block fading model, where $\hij$ remains constant over one block, and varies independently from a block to another according to a Nakagami-$m$ fading model with shape parameter $\mij$ and average power $\mathbb{E}\left\{\gij\right\}=\piij$. Accordingly, $\gij$ is a Gamma-distributed \ac{RV} with shape parameter $\mij$ and scale parameter $\thetaij=\frac{\piij}{\mij}$, for which we use the shorthand notation $\gij \sim \GAMMA\left(\mij,\thetaij\right)$. For a Gamma \ac{RV} $\RV \sim \GAMMA\left(m,\theta\right)$, the \ac{PDF} and the \ac{CDF} are given by
\begin{eqnarray}
\!\!\!f_{\RV}\left(\RVval;m,\theta\right) = \frac{\RVval^{m-1}e^{-\frac{\RVval}{\theta}}}{\GAMFN{m}\theta^m} \text{ and }
F_{\RV}\left(\RVval;m,\theta\right) = \frac{\LWRGAMFN{m,\frac{\RVval}{\theta}}}{\GAMFN{m}}
\label{eq:gamma_cdf},\!\!\!\!\!\!
\end{eqnarray}
\noindent respectively, where $\LWRGAMFN{a,b}=\int_0^b t^{a-1}e^{-t}dt$ denotes the lower incomplete Gamma function, and $\GAMFN{a}$ denotes the Gamma function \cite{abramowitz_stegun}. All channel fading gains are assumed to be mutually independent. The source and the $k^{\rm th}$ relay powers are denoted by $\PS$ and $\Pk$, respectively. Also, $\Nk$ and $\ND$ denote the complex \ac{AWGN} components at the $k^{\rm th}$ relay and the destination, with variance $\nvark$ and $\nvarD$, respectively.

As commonly assumed in the literature, for instance in \cite{201102_TWC_Lee_DF}, we consider the relays to be clustered somewhere between the source and the destination. Hence, the distances among the relays are much shorter than those between the relays and the source/destination.
Thus, it is reasonable to assume the symmetric scenario where all source-relay links have an average gain of $\mathbb{E}\{\gsk\}=\pisr$, while all relay-destination links have an average gain of $\mathbb{E}\{\gkd\}=\pird$, $\forall k \in \KSet$. Also, we assume that all relays have their \ac{RSI} links with the same average gain, i.e., $\mathbb{E}\{\gkk\}=\pirr$. All relays have the same average gain to the primary receiver, denoted as $\pirp$. Further, it is assumed that $\msk=\msr$, $\mkd=\mrd$, $\mkk=\mrr$ and $\mkp=\mrp$, for all $k\in\KSet$. Although the analysis of asymmetric scenarios remains possible, the previous assumptions allow for simpler final expressions, and yet maintain the same diversity order of the system. Therefore, it follows that $\thetask=\thetasr$, $\thetakd=\thetard$, $\thetakk=\thetarr$ and $\thetakp=\thetarp$ for all $k\in\KSet$. Finally, we assume that $\nvark=\nvarD=1$, while all relays have the same transmit power $\Pk=\PR$.

\subsection{Signal Model}
When the $k^{\rm th}$ relay is selected, the received signals at the $k^{\rm th}$ relay and destination at time $\TM$ are given, respectively, by
\begin{eqnarray}
\Yk[\TM]	
&=& \sqrt{\PS} \hsk \XS[\TM] + \sqrt{\PR^\PWRSCL} \hkk \Xk[\TM] + \Nk[\TM], \label{eq:R_RX_FDRS_inst}\\
\YD[\TM]	
&=& \sqrt{\PR} \hkd \Xk[\TM] + \sqrt{\PS} \hsd \XS[\TM] +\ND[\TM], \label{eq:D_RX_FDRS_inst}
\end{eqnarray}
where $\PI$ and $\XI[\TM]$ denote the transmit power and the transmit symbol at time $t$ at node $\iN \in \{\SN, \RN\}$, respectively, while $\NI[\TM]$ represents the \ac{AWGN} component at node $\iN \in \{\RN, \DN\}$ at time $\TM$. It is assumed that the \ac{RSI} term proportionally scales with $\PR^\PWRSCL$, where $0 \leq \PWRSCL \leq 1$, covering the range from constant to linear scaling with the relay power. Without loss of generality, all \ac{AWGN} components are assumed of unit variance.

Due to the source and relay asynchronous transmissions, the signal transmitted by the relay (source) is considered in this class as an additional noise term at the relay (destination) as commonly treated in the related literature \cite{riihonen201109TWC}.  Accordingly, the received \acp{SINR} at the $k^{\rm th}$ relay and at the destination are given respectively by
\begin{eqnarray}
\SkFDSNR=\frac{\PS\gsk}{\PR^\PWRSCL\gkk + 1}\,\,\,\, \text{and }
\kDFDSNR=\frac{\PR\gkd}{\PS\gsd + 1}. \label{SNRs_of_hops_FDRS}
\end{eqnarray}
Since only the information via the multi-hop path is leveraged, this protocol is referred to in the recent literature as the \ac{MHDF} protocol.
\break\indent\textit{Underlay Cognitive Setting: }
In an underlay setting, the introduced interference level on node $\PN$ is constrained not to exceed a certain threshold, $\Ith$. Thus, when the $k^{\rm th}$ relay is selected, the interference constraint is given by
\begin{eqnarray}
\Isk = \Isp + \Ikp \leq \Ith, \label{interference_constraint_NDL}
\end{eqnarray}
where $\Isp =\PS \gsp$ and $\Ikp = \PR \gkp$ are the interference components imposed on the primary receiver due to the source and the $k^{\rm th}$ relay, respectively, while $\Isk$ denotes their sum. When no relays are selected while a direct source-destination link exists, the interference constraint is simply given by
\begin{equation}
\Isp \leq \Ith.
\end{equation}
\section{Full-Duplex Relay Selection}
We first derive the end-to-end \ac{SINR} \ac{CDF} for opportunistic \ac{MHDF} \ac{FDRS} with/without a direct ${\rm S - D}$ link. Specifically, when $K$ \ac{DF} full-duplex relays are available, the end-to-end \ac{SINR} is given by
\begin{eqnarray}
\etoeFDSNR
&=& \max_{k\in \KSet}\left\{\kFDSNR\right\}, \label{e2e_snr_NDL_DL}
\end{eqnarray}
where
\begin{eqnarray}
\kFDSNR
&=& \min\left\{\SkFDSNR,\kDFDSNR\right\}. \label{k_th_hop_SINR}
\end{eqnarray}
In the absence of a direct link, $\kDFDSNR$ is calculated for $\SDFDSNR=\PS\gsd=0$. The expression in \eqref{e2e_snr_NDL_DL} also applies to the scenario where the direct link exists, yet it is treated as mere interference. Alternatively, if \ac{DT} is taken into account as a possible diversity branch, the end-to-end \ac{SINR} is given by
\begin{eqnarray}
\etoeFDSNR
&=& \max\left\{\max_{k\in \KSet}\left\{\kFDSNR\right\},\SDFDSNR\right\}. \label{general_e2e_snr_hybrid}
\end{eqnarray}
In what follows, we derive the exact end-to-end \ac{SINR} \ac{CDF} for three \ac{MHDF} scenarios; 1) \ac{NDL}, 2) \ac{IDL} and 3) hybrid \ac{IDL}/\ac{DT}.

\subsection{CDF of Link SINRs}
\subsubsection{First Hop} According to the channel model explained in Section \ref{subsec:chan_model}, the \ac{CDF} of the \ac{SINR} pertaining to the first hop of the $k^{th}$ path is given as
\begin{eqnarray}
F_{\SkFDSNR}\left(\RVval\right) = F_{Z}\left(\RVval;\pv_1\right),
\end{eqnarray}
where $\pv_1=\left(\msr,\PS\thetasr,\mrr,\PR^\PWRSCL\thetarr\right)$, while $F_{Z}\left(\RVval;\pv\right)$ is given in the following remark \cite{FDR_Nakagami_NCC_paper,201303_WCOML_Alves}.
\begin{remark}[CDF of $Z=\frac{X_1}{X_2+1}$]\label{theorem_gamma_RV_ratio_CDF}
The \ac{CDF} of $Z=\frac{X_1}{X_2+1}$, where $X_i \sim \GAMMA\left(m_{i},\theta_{i}\right)$, for $i \in \{1,2\}$, are \ac{inid} RVs,  for general real-valued $m_1\geq \frac{1}{2}$ and integer-valued $m_2\geq 1$, is given by \cite{FDR_Nakagami_NCC_paper,201303_WCOML_Alves}:
\begin{equation}
F_{Z}\left(z;\pv\right)=\frac{\LWRGAMFN{m_1,\frac{z}{\theta_1}}}{\GAMFN{m_1}}+B\sum_{k=0}^{m_2-1}\frac{c^{-d}}{{\theta_2}^{k}} \WTKRFN{a}{b}{c}, \label{CDF_ratio_two_gamma_RVs}
\end{equation}
where $\pv=\left(m_1,\theta_1,m_2,\theta_2\right)$ is a vector of distribution parameters, $\LWRGAMFN{\alpha,\beta}=\int_0^\beta t^{\alpha-1}e^{-t}dt$ is the lower incomplete Gamma function, $\WTKRFN{a}{b}{c}$ is the Whittaker function \cite[Eq. 13.1.33]{abramowitz_stegun}, $a=\frac{m_1-k-1}{2}$, $b=\frac{-m_1-k}{2}$, $c=\frac{z}{\theta_1}+\frac{1}{\theta_2}$, $d=\frac{m_1+k+1}{2}$ and
\begin{eqnarray}
B&=&\frac{\exp\left({-\frac{1}{2}\left(\frac{z}{\theta_1}-\frac{1}{\theta_2}\right)}\right)}{\GAMFN{m_1}}\left(\frac{z}{\theta_1}\right)^{m_1}.
\end{eqnarray}
\end{remark}
\subsubsection{Second Hop} Since the direct link \ac{SNR}, $\SDFDSNR$, is a common \ac{RV} among all the multi-hop paths, all the second-hop gains are clearly correlated. However, they are conditionally independent given $\SDFDSNR=\beta$. Thus,  in the presence of a direct link, we are only interested in the conditional distributions of the second-hop gains given $\SDFDSNR=\beta$, which follow $\kDFDSNR  \vert  \SDFDSNR \sim \GAMMA\left(\mrd, \frac{\PR\thetard}{\beta+1}\right)$, $\forall k \in \mathcal{K}$. On the other hand, when no direct link exists, it is clear that $\kDFDSNR  \vert  \SDFDSNR \sim \GAMMA\left(\mrd, \PR\thetard\right)$.

\subsubsection{The $k^{th}$ Multi-Hop Path} According to \eqref{k_th_hop_SINR} and with the above-mentioned distributions of the hop \acp{SINR}, the conditional \ac{CDF} of the \ac{SINR} over the $k^{th}$ path given the direct link \ac{SNR} $\SDFDSNR=\beta$ is given by
\begin{eqnarray}
F_{\kFDSNR \vert \SDFDSNR}(\RVval \vert \beta)
=1-\overline{F_{Z}}\left(\RVval;\pv_1\right) \overline{F_X}\left(\RVval;\mrd,\frac{\PR\thetard}{\beta+1}\right),
\end{eqnarray}
where $\overline{F}(\cdot)=1- F(\cdot)$ denotes the complementary \ac{CDF}. It is clear that in the absence of a direct link, the \ac{CDF} of the $k^{th}$ path \ac{SINR} is given by
\begin{eqnarray}
F_{\kFDSNR}(\RVval)
=1-\overline{F_{Z}}\left(\RVval;\pv_1\right) \overline{F_X}\left(\RVval;\mrd,\PR\thetard\right).
\end{eqnarray}
\subsection{CDF of End-to-End SINR}
\subsubsection{\ac{NDL}} In the absence of a direct link, the end-to-end \ac{SINR} \ac{CDF} has the following simple form:
\begin{eqnarray}
\!\!\!\!\!\!\!\!\!\!\CDFNDL
\!\!&=&\!\!  \left(  1-\overline{F_{Z}}\left(\RVval;\pv_1\right) \overline{F_X}\left(\RVval;\mrd,\PR\thetard\right) \right)^K,
\end{eqnarray}
which is explicitly given in \eqref{CDF_NDL} for real-valued $\msr$ and $\mrd$, and integer-valued $\mrr$.
\begin{figure*}
\centering
\begin{equation}
\begin{split}
\CDFNDL
&=\Bigg(1-\frac{\UPRGAMFN{\mrd,\frac{\RVval}{\PR\thetard}}}{\GAMFN{\msr}\GAMFN{\mrd}}\Bigg(\UPRGAMFN{\msr,\frac{\RVval}{\PS\thetasr}} - e^{-\frac{1}{2}\left(\frac{\RVval}{\PS\thetasr}-\frac{1}{\PR^\PWRSCL\thetarr}\right)}\left(\frac{\RVval}{\PS\thetasr}\right)^{\msr}\\
&\times \sum_{l=0}^{\mrr-1}\frac{\left(\frac{\RVval}{\PS\thetasr}+\frac{1}{\PR^\PWRSCL\thetarr}\right)^{-\frac{\msr+l+1}{2}}}{{\left(\PR^\PWRSCL\thetarr\right)}^{l}} \WTKRFN{\frac{\msr-l-1}{2}}{\frac{-\msr-l}{2}}{\frac{\RVval}{\PS\thetasr}+\frac{1}{\PR^\PWRSCL\thetarr}}\Bigg)\Bigg)^K.
\end{split}\label{CDF_NDL}
\end{equation}
\hrule
\end{figure*}
\subsubsection{\ac{IDL}}
When a direct link exists, yet treated as interference to the multi-hop paths, the end-to-end \ac{SINR} \ac{CDF} is given by
\begin{eqnarray}
\CDFIDL
&=&\int_0^\infty F_{\kFDSNR \vert \SDFDSNR}(\RVval \vert \beta)^K f_x\left(\beta;\msd,\PS\thetasd\right) \mathrm{d}\beta\nonumber\\
&=&\sum_{k=0}^K C(\RVval,k)~~\mathcal{I}_1,
\end{eqnarray}
with \vspace{-4mm}
\begin{eqnarray}
\mathcal{I}_1
&=&
\int_0^\infty \frac{\UPRGAMFN{\mrd,\frac{\RVval(\beta+1)}{\PR\thetard}}^k}{\GAMFN{\mrd}^k} \beta^{\msd-1}e^{-\frac{\beta}{\PS\thetasd}} \mathrm{d}\beta,\nonumber\\
C(\RVval,k)
&=&
{K \choose k} \frac{\left(-\overline{F_{Z}}\left(\RVval;\pv_1\right) \right)^k}{\GAMFN{\msd}{\left(\PS\thetasd\right)}^{\msd}}\nonumber,
\end{eqnarray}
where the binomial expansion is utilized in the last step. We are now interested in solving the integral $\mathcal{I}_1$. Using the finite series expansion of the upper incomplete Gamma function in \cite[Eq. 8.352-2]{gradshteyn_ryzhik}, for integer values of $\mrd$, we get
\begin{eqnarray}
\!\mathcal{I}_1
\!\!\!\!\!&=&\!\!\!\!\! \int_0^\infty \!\! \left(e^{-\frac{\RVval(\beta+1)}{\PR\thetard}}\sum_{n=1}^{\mrd}  \frac{\left(\frac{\RVval(\beta+1)}{\PR\thetard}\right)^{n-1}}{\GAMFN{n}}  \right)^k    \beta^{\msd-1}e^{-\frac{\beta}{\PS\thetasd}} \mathrm{d}\beta\nonumber\\
\!\!\!\!&=&\!\!\!\! e^{-\frac{\RVval k}{\PR\thetard}} \sum_{\sum_{n=1}^{\mrd} k_{n} = k} C_{\{k_n\}} \mathcal{I}_2,
\label{multinomial_theorem_applied_1}
\end{eqnarray}
where \vspace{-4mm}
\begin{eqnarray}
\mathcal{I}_2&=&\int_0^\infty (\beta+1)^{D_{\{k_n\}}}\beta^{\msd-1}e^{-\beta\eta_k} \mathrm{d}\beta,\label{I_2}\\
\eta_k&=&\left(\frac{1}{\PS\thetasd}+\frac{\RVval k}{\PR\thetard}\right),\\
C_{\{k_n\}}&=&\frac{\GAMFN{k+1}\left(\frac{\RVval}{\PR\thetard}\right)^{\sum_{n=1}^{\mrd}k_n(n-1)}}{\prod_{n=1}^{\mrd}\left(\GAMFN{k_n+1}\GAMFN{n}^{k_n}\right)},\\
D_{\{k_n\}}&=&\sum_{n=1}^{\mrd}k_n(n-1),
\end{eqnarray}
with \eqref{multinomial_theorem_applied_1} obtained via the multinomial theorem \cite[Section 24.1.2]{abramowitz_stegun}. By substitution of variables, $y=\beta+1$, $\mathcal{I}_2$ in \eqref{I_2} is the Riemann-Liouville integral in \cite[Eq. 3.383-4]{gradshteyn_ryzhik}. Hence,
\begin{eqnarray}
\mathcal{I}_2
&=&e^{\eta_k}\int_1^\infty y^{D_{\{k_n\}}}(y-1)^{\msd-1}e^{-y\eta_k} \mathrm{d}y\nonumber\\
&=&e^{\frac{\eta_k}{2}}\eta_k^{-\frac{\msd+D_{\{k_n\}}+1}{2}}\GAMFN{\msd}\nonumber\\
&& \times \WTKRFN{\frac{D_{\{k_n\}} -\msd+1}{2}}{-\frac{\msd+D_{\{k_n\}}}{2}}{\eta_k}.
\end{eqnarray}
Accordingly, $\CDFIDL$ is given by \eqref{CDF_IDL} for integer-valued $\mrr$ and $\mrd$, and real-valued $\msr$ and $\msd$.
\begin{figure*}
\centering
\begin{footnotesize}
\begin{equation}
\begin{split}
\CDFIDL
&=\sum_{k=0}^K {K \choose k} \frac{e^{\left(\frac{1}{2\PS\thetasd}-\frac{\RVval k}{2\PR\thetard}\right)}}{\left(\PS\thetasd\right)^{\msd}\GAMFN{\msr}^k}
\Bigg(e^{-\frac{1}{2}\left(\frac{\RVval}{\PS\thetasr}-\frac{1}{\PR^\PWRSCL\thetarr}\right)}\left(\frac{\RVval}{\PS\thetasr}\right)^{\msr}
\\&\times
{\displaystyle\sum_{l=0}^{\mrr-1}} \frac{\left(\frac{\RVval}{\PS\thetasr}+\frac{1}{\PR^\PWRSCL\thetarr}\right)^{-\frac{\msr+l+1}{2}}}{\left(\PR^\PWRSCL\thetarr\right)^{l}}
\WTKRFN{\frac{\msr-l-1}{2}}{\frac{-\msr-l}{2}}{\frac{\RVval}{\PS\thetasr}+\frac{1}{\PR^\PWRSCL\thetarr}}        - \UPRGAMFN{\msr,\frac{\RVval}{\PS\thetasr}}
\Bigg)^k\\
&\times \sum_{\sum_{n=1}^{\mrd} k_{n} = k} \frac{\GAMFN{k+1}\left(\frac{\RVval}{\PR\thetard}\right)^{\sum_{n=1}^{\mrd}k_n(n-1)}}{\prod_{n=1}^{\mrd}\left(\GAMFN{k_n+1}\GAMFN{n}^{k_n}\right)} \eta_k^{-\frac{\msd+\sum_{n=1}^{\mrd}k_n(n-1)+1}{2}}
\WTKRFN{\frac{\sum_{n=1}^{\mrd}k_n(n-1) -\msd+1}{2}}{-\frac{\msd+\sum_{n=1}^{\mrd}k_n(n-1)}{2}}{\eta_k}.
\end{split}
\label{CDF_IDL}
\end{equation}
\end{footnotesize}
\hrule
\end{figure*}
\subsubsection{Hybrid \ac{IDL}/\ac{DT}}
When the $\SN - \DN$ link is leveraged as an additional diversity path, $\SDFDSNR=\beta$ is bounded above by $\RVval$. Hence, the \ac{CDF} expression is similar to that of the \ac{IDL} case, however with the upper integration limit changed to $\RVval$ to be
\begin{eqnarray}
\CDFMHDT
=\int_0^\RVval F_{\kFDSNR \vert \SDFDSNR}(\RVval \vert \beta)^K f_x\left(\beta;\msd,\PS\thetasd\right) \mathrm{d}\beta.
\end{eqnarray}
The evaluation of such an integral follows the same steps, except that the integral $\mathcal{I}_2$ is changed to
\begin{eqnarray}
\tilde{\mathcal{I}}_2
\!\!\!\!&=&\!\!\!\!\int_0^{\RVval} (\beta+1)^{D_{\{k_n\}}}\beta^{\msd-1}e^{-\beta\eta_k} \mathrm{d}\beta\nonumber\\
\!\!\!\!&=&\!\!\!\! \sum_{r=0}^{D_{\{k_n\}}} {D_{\{k_n\}} \choose r}\int_0^{\RVval}\beta^{r+\msd-1}e^{-\beta\eta_k} \mathrm{d}\beta\nonumber\\
\!\!\!\!&=&\!\!\!\! \sum_{r=0}^{D_{\{k_n\}}} {D_{\{k_n\}} \choose r} \eta_k^{-(r+\msd)} \LWRGAMFN{r+\msd, \RVval\eta_k},
\end{eqnarray}
for integer-valued $\msd$ due to the use of the binomial theorem, with the last integral evaluated using \cite[Eq. 3.381-1]{gradshteyn_ryzhik}. Hence, $\CDFMHDT$ is finally given by \eqref{CDF_MHDT} for real-valued $\msr$ and integer-valued $\mrr$, $\mrd$ and $\msd$.

\begin{figure*}
\centering
\begin{footnotesize}
\begin{equation}
\begin{split}
\CDFMHDT
&=\sum_{k=0}^K {K \choose k} \frac{e^{-\frac{\RVval k}{\PR\thetard}}}{\GAMFN{\msd}\left(\PS\thetasd\right)^{\msd}\GAMFN{\msr}^k} \Bigg(e^{-\frac{1}{2}\left(\frac{\RVval}{\PS\thetasr}-\frac{1}{\PR^\PWRSCL\thetarr}\right)}\left(\frac{\RVval}{\PS\thetasr}\right)^{\msr}\\
& \times {\displaystyle\sum_{l=0}^{\mrr-1}} \frac{\left(\frac{\RVval}{\PS\thetasr}+\frac{1}{\PR^\PWRSCL\thetarr}\right)^{-\frac{\msr+l+1}{2}}}{\left(\PR^\PWRSCL\thetarr\right)^{l}}
\WTKRFN{\frac{\msr-l-1}{2}}{\frac{-\msr-l}{2}}{\frac{\RVval}{\PS\thetasr}+\frac{1}{\PR^\PWRSCL\thetarr}}        - \UPRGAMFN{\msr,\frac{\RVval}{\PS\thetasr}}  \Bigg)^k\\
&\times \sum_{\sum_{n=1}^{\mrd} k_{n} = k} \frac{\GAMFN{k+1}\left(\frac{\RVval}{\PR\thetard}\right)^{\sum_{n=1}^{\mrd}k_n(n-1)}}{\prod_{n=1}^{\mrd}\left(\GAMFN{k_n+1}\GAMFN{n}^{k_n}\right)}
\sum_{r=0}^{\sum_{n=1}^{\mrd}k_n(n-1)} {\sum_{n=1}^{\mrd}k_n(n-1) \choose r} \eta_k^{-(r+\msd)} \LWRGAMFN{r+\msd, \RVval\eta_k}.
\end{split}
\label{CDF_MHDT}
\end{equation}
\end{footnotesize}
\hrule
\vspace{-2mm}
\end{figure*}

\subsubsection{Selective decode-and-forward}
Unlike the previous non-selective multi-hop protocols, selective block full-duplex cooperation is recently studied in \cite{6510556,2015XX_TWC_Khafagy_FDR}. In this class, both the multi-hop path and the direct path are simultaneously exploited as two information-bearing paths. In order to leverage the information in the two paths in the presence of their relative delay, a block transmission scheme is adopted \cite{6510556}. When \ac{SDF} is considered, the instantaneous \ac{SINR} in the first hop via the $k^{th}$ relay, its virtual \ac{MISO} channel with the source and the $k^{th}$ relay as the transmitter side, and the direct link are given, respectively, by
\begin{eqnarray}
\SkFDSNR				&=&	\frac{\PS\gsk}{\PR^\PWRSCL\gkk+1},\\
\SkDFDSNR 				&\approx&	\PR\gkd+\PS\gsd,\\
\SDFDSNR				&=&	\PS\gsd.
\end{eqnarray}
The relay selection protocol is assumed to adopt an incremental relaying fashion, where the \ac{DT} is favored as long as it can by itself support the end-to-end communication. Otherwise, one of the relays assists. Seeking the highest instantaneous \ac{SNR} when opportunistic relay selection is adopted, the end-to-end \ac{SNR} via the best relay is accordingly given by
\begin{eqnarray}
\etoeFDSNR = \max\left\{\max_{k\in \KSet}\left\{\min\left\{\SkFDSNR	,\SkDFDSNR\right\}\right\},\SDFDSNR\right\}.
\end{eqnarray}
From the link gains independence, we notice that the \acp{SNR} of all $K$ paths are conditionally independent given the direct link \ac{SNR} $\SDFDSNR$. Thus, the end-to-end \ac{SNR} \ac{CDF} is given by
\begin{eqnarray}
F_{\etoeFDSNR}^{\rm SDF}(\RVval)=\int_0^\RVval \left(F_{\kFDSNR \vert \SDFDSNR}^{\rm SDF}(\RVval \vert \beta)\right)^K f_{\SDFDSNR}(\beta)\mathrm{d}\beta,
\end{eqnarray}
where $F_{\kFDSNR \vert \SDFDSNR}^{\rm SDF}(\RVval \vert \beta)$ is raised to the $K^{th}$ power due to the \ac{iid} link gains.

The conditional \ac{CDF} of the $k^{th}$ path given the direct link gain is given, for $0 \leq \beta\leq \RVval$, by
\begin{eqnarray}
F_{\kFDSNR \vert \SDFDSNR}^{\rm SDF}(\RVval \vert \beta)
&=& 1-\overline{F_{\SkFDSNR \vert \SDFDSNR}^{\rm SDF}}(\RVval \vert \beta) \overline{F_{\SkDFDSNR \vert \SDFDSNR}^{\rm SDF}}(\RVval \vert \beta)\nonumber\\
&\approx &
1-\overline{F_{Z}}\left(\RVval;\pv_1\right)\overline{F_\RV}\left(\RVval-\beta;\mrd,\PR\thetard\right), \nonumber
\end{eqnarray}
\noindent since $\SkDFDSNR \approx \RDFDSNR + \SDFDSNR \geq \SDFDSNR$. Hence,
\begin{equation}
\begin{split}
\!\!\!F_{\etoeFDSNR}^{\rm SDF}(\RVval)
&\approx
\int_0^\RVval \Bigg(
1-\overline{F_{Z}}\left(\RVval;\pv_1\right) \frac{\UPRGAMFN{\mrd,\frac{\RVval-\beta}{\PR\thetard}}}{\GAMFN{\mrd}}
\Bigg)^K
\\& \times
\frac{\beta^{\msd-1}e^{-\frac{\beta}{\PS\thetasd}}}{\GAMFN{\msd}\left(\PS\thetasd\right)^\msd} \mathrm{d}\beta
\end{split}
\end{equation}
\begin{equation}
\begin{split}
&=\sum_{k=0}^K {K \choose k} \Bigg(
-\overline{F_{Z}}\left(\RVval;\pv_1\right)\Bigg)^k \frac{1}{\GAMFN{\msd}\left(\PS\thetasd\right)^\msd}
\\
& \times \underbrace{\int_0^\RVval\Bigg(\frac{\UPRGAMFN{\mrd,\frac{\RVval-\beta}{\PR\thetard}}}{\GAMFN{\mrd}}
\Bigg)^k
\beta^{\msd-1}e^{-\frac{\beta}{\PS\thetasd}} \mathrm{d}\beta}_{\mathcal{{\hat{I}}}_1},
\end{split}
\label{end_to_end_SNR_CDF_SDF_given_L_Nakagami_step}
\end{equation}
where the binomial theorem is utilized in the last step. For integer $\mrd$, we can use the following series expansion for the upper regularized Gamma function \cite[Eq. 8.352-2]{gradshteyn_ryzhik}:
\begin{equation}
\frac{\UPRGAMFN{\mrd,\frac{\RVval-\beta}{\PR\thetard}}}{\GAMFN{\mrd}}
=
e^{-\frac{\RVval-\beta}{\PR\thetard}}\sum_{m=0}^{\mrd-1}\frac{\left(\frac{\RVval-\beta}{\PR\thetard}\right)^m}{\GAMFN{m+1}}.\label{series_expan_upper_regularized_gamma_integer}
\end{equation}
Hence,
\begin{eqnarray}
\mathcal{\hat{I}}_1
&=& e^{-\frac{k\RVval}{\PR\thetard}} \sum_{\sum_{n=1}^{\mrd} \hat{k}_{n} = k} \hat{C}_{\{\hat{k}_n\}} \mathcal{\hat{I}}_2,\label{multinomial_theorem_applied_SDF}
\end{eqnarray}
where
\begin{eqnarray}
\mathcal{\hat{I}}_2&=&\int_0^\RVval (\RVval-\beta)^{\hat{D}_{\{\hat{k}_n\}}}\beta^{\msd-1}e^{-\beta\hat{\eta}_k} \mathrm{d}\beta,\label{I_2_SDF}\\
\hat{\eta}_k&=&\left(\frac{1}{\PS\thetasd}-\frac{k}{\PR\thetard}\right),\\
\hat{C}_{\{\hat{k}_n\}}&=&\frac{\GAMFN{k+1}\left(\frac{1}{\PR\thetard}\right)^{\sum_{n=1}^{\mrd}\hat{k}_n(n-1)}}{\prod_{n=1}^{\mrd}\left(\GAMFN{\hat{k}_n+1}\GAMFN{n}^{\hat{k}_n}\right)},\\
\hat{D}_{\{\hat{k}_n\}}&=&\sum_{n=1}^{\mrd}\hat{k}_n(n-1),
\end{eqnarray}
with \eqref{multinomial_theorem_applied_SDF} obtained via the multinomial theorem. Now, the integral $\mathcal{\hat{I}}_2$ in \eqref{I_2_SDF} is on the form of the Riemann-Liouville integral in \cite[Eq. 3.383-1]{gradshteyn_ryzhik} when $\hat{\eta}_k \leq 0$. Hence,
\begin{eqnarray}
\!\!\!\!\!\!\!\!\mathcal{\hat{I}}_2
&=& \RVval^{\hat{D}_{\{\hat{k}_n\}}+\msd} \BTAFN{\hat{D}_{\{\hat{k}_n\}}+1}{\msd}
\nonumber\\&&\times
\KMRFN{\msd}{\hat{D}_{\{\hat{k}_n\}}+\msd+1}{-\RVval \hat{\eta}_k}.
\end{eqnarray}
On the other hand, when $\hat{\eta}_k > 0$, we can use the change of variables $y=\RVval-\beta$ and again use \cite[Eq. 3.383-1]{gradshteyn_ryzhik} to get
\begin{eqnarray}
\!\!\!\!\!\!\!\!\mathcal{\hat{I}}_2
&=& e^{-\RVval \hat{\eta}_k} \RVval^{\hat{D}_{\{\hat{k}_n\}}+\msd} \BTAFN{\msd}{\hat{D}_{\{\hat{k}_n\}}+1}\nonumber\\&&\times
\KMRFN{\hat{D}_{\{\hat{k}_n\}}+1}{\hat{D}_{\{\hat{k}_n\}}+\msd+1}{\RVval \hat{\eta}_k}.
\end{eqnarray}
Finally, substituting back, we get the end-to-end \ac{SNR} \ac{CDF} shown in \eqref{end_to_end_SNR_CDF_SDF_given_L_Nakagami}.
\begin{figure*}
\centering
\begin{eqnarray}
\!\!\!F_{\etoeFDSNR}^{\rm SDF}(\RVval)
\!\!\!&\approx&\!\!\!\sum_{k=0}^K {K \choose k} \Bigg(
-\frac{\UPRGAMFN{\msr,\frac{\RVval}{\PS\thetasr}}}{\GAMFN{\msr}}+
\frac{\exp\left({-\frac{1}{2}\left(\frac{\RVval}{\PS\thetasr}-\frac{1}{\PR^\PWRSCL\thetarr}\right)}\right)}{\GAMFN{\msr}}\left(\frac{\RVval}{\PS\thetasr}\right)^{\msr}
\nonumber\\
&& \times \sum_{l=0}^{\mrr-1}\frac{\left(\frac{\RVval}{\PS\thetasr}+\frac{1}{\PR^\PWRSCL\thetarr}\right)^{-\frac{\msr+l+1}{2}}}{\left(\PR^\PWRSCL\thetarr\right)^{l}} \WTKRFN{\frac{\msr-l-1}{2}}{\frac{-\msr-l}{2}}{\frac{\RVval}{\PS\thetasr}+\frac{1}{\PR^\PWRSCL\thetarr}}\Bigg)^k
\nonumber\\
&& \times \frac{1}{\GAMFN{\msd}\left(\PS\thetasd\right)^\msd} e^{-\frac{k\RVval}{\PR\thetard}} \sum_{\sum_{n=1}^{\mrd} \hat{k}_{n} = k} \hat{C}_{\{\hat{k}_n\}} \mathcal{\hat{I}}_2.\label{end_to_end_SNR_CDF_SDF_given_L_Nakagami}
\end{eqnarray}
\hrule
\end{figure*}
\section{Underlay Cognitive Relay Selection}
There exists an interference threshold, $\Ith$, that the secondary system cannot exceed. Now, we are interested in the probability to have $L$ out of the $K$ available relays that satisfy the interference constraint $\Ith$. Let us denote this probability by $\PL$. We first start with the scenarios where the direct link is not leveraged.
\subsection{No Direct $\SN - \DN$ Link}
When no $\SN - \DN$ link exists,  the end-to-end \ac{SNR} \ac{CDF} is given as
\begin{eqnarray}
F^{\rm NDL}_{\etoeFDSNR}(\RVval)
&=&\sum_{L=0}^K F^{\rm NDL}_{\etoeFDSNR}(\RVval \vert L) \PL,
\end{eqnarray}
where $F^{\rm NDL}_{\etoeFDSNR}(\RVval \vert L)$ is that derived in the previous section when $L$ relays are available. The same expression also applies to the \ac{IDL} scenario where the direct link is not leveraged. Specifically,
\begin{eqnarray}
F^{\rm IDL}_{\etoeFDSNR}(\RVval)
&=&\sum_{L=0}^K F^{\rm IDL}_{\etoeFDSNR}(\RVval \vert L) \PL.
\end{eqnarray}

Now, we derive the probability of $L$ feasible relays, $\PL$. When the $k^{\rm th}$ relay is active, the interference constraint is given by \eqref{interference_constraint_NDL}. Unlike half-duplex cognitive relay settings, full-duplex operation causes a superposition of the source and relay interference at the primary receiver. Since the source interference is a common random variable when any of the $K$ relays is selected, the superimposed interference signals are correlated. However, they are conditionally independent given $\Isp$. Conditioned on $\Isp=\beta\geq 0$, the probability that the $k^{\rm th}$ relay is feasible is given by
\begin{equation}
F_{\Isk \vert \Isp}(\Ith \vert \beta)
=\begin{cases}
F_\RV\left(\Ith-\beta;\mrp,\PR\thetarp\right)
, &\mbox{if }  \Ith>\beta,\\ 0, & \mbox{elsewhere.}\end{cases}\label{conditional_distribution_of_interference_paths}
\end{equation}
Accordingly, conditioned on $\Isp=\beta$ for $0 \leq \beta \leq \Ith$, and due to the considered symmetric scenario, $\mathcal{P}_{L \vert \Isp}(\beta)$ is given in terms of a binomial distribution as
\begin{equation}
\!\!\!\!\!\!\mathcal{P}_{L \vert \Isp}(\beta)
\!=\! {K \choose L} \sum_{l=0}^L {L \choose l}(-1)^l \left(\frac{\UPRGAMFN{\mrp,\frac{\Ith-\beta}{\PR\thetarp}}}{\GAMFN{\mrp}}\right)^{\!\!\!\!K-L+l}\!\!\!\!\!\!\!\!\!\!\!\!\!,\label{conditional_PL_Nakagami}
\end{equation}
where the binomial expansion is again utilized. It is straightforward to see that when $\beta>\Ith$, $\mathcal{P}_{L \vert \Isp}(\beta)=0$ for $L=1,2,\cdots,K$, while $\mathcal{P}_{L \vert \Isp}(\beta)=1$ for $L=0$. Hence, for the case when $L=0$, the support of $\beta$ is $0\leq\beta\leq\infty$, while it is $0\leq\beta\leq \Ith$ for $1 \leq L \leq K$. Now, we can obtain $\PL$ for $L=1,2,\cdots,K$ as
\begin{eqnarray}
\!\!\!\!\!\!\!\!\PL\!\!\!\!
&=&
\!\!\!\! {K \choose L} \sum_{l=0}^L  \frac{{L \choose l}(-1)^l}{\GAMFN{\msp}\left(\PS\thetasp\right)^\msp} \mathcal{\check{I}}_1,
\end{eqnarray}
where
\begin{eqnarray}
\mathcal{\check{I}}_1
=
\int_0^{\Ith} \left(\frac{\UPRGAMFN{\mrp,\frac{\Ith-\beta}{\PR\thetarp}}}{\GAMFN{\mrp}}\right)^{K-L+l} \!\!\!\!\beta^{\msp-1}e^{-\frac{\beta}{\PS\thetasp}} \mathrm{d}\beta.\nonumber
\end{eqnarray}
The integral in the last step can be evaluated with the aid of the multinomial theorem expansion as previously done in \eqref{end_to_end_SNR_CDF_SDF_given_L_Nakagami_step}. Specifically, it is given by
\begin{eqnarray}
\!\!\!\!\!\mathcal{\check{I}}_1
&=& e^{-\frac{(K-L+l)\Ith}{\PR\thetarp}} \sum_{\sum_{n=1}^{\mrp} \check{k}_{n} = K-L+l} \check{C}_{\{\check{k}_n\}} \mathcal{\check{I}}_2,\label{multinomial_theorem_applied_SDF}
\end{eqnarray}
where
\begin{eqnarray}
\!\!\!\!\!\mathcal{\check{I}}_2
\!\!\!&=&\!\!\!
\int_0^{\Ith} (\Ith-\beta)^{\check{D}_{\{\check{k}_n\}}}\beta^{\msp-1}e^{-\beta\check{\eta}_l} \mathrm{d}\beta,\label{I_2_CogFDRS_NDL}\\
\!\!\!\!\!\check{\eta}_l
\!\!\!&=&\!\!\!
\left(\frac{1}{\PS\thetasp}-\frac{K-L+l}{\PR\thetarp}\right),
\end{eqnarray}
\begin{eqnarray}
\!\!\!\!\!\check{C}_{\{\check{k}_n\}}\!\!\!&=&\!\!\!\frac{\GAMFN{K-L+l+1}\left(\frac{1}{\PR\thetarp}\right)^{\sum_{n=1}^{\mrp}\check{k}_n(n-1)}}{\prod_{n=1}^{\mrp}\left(\GAMFN{\check{k}_n+1}\GAMFN{n}^{\check{k}_n}\right)},\\
\!\!\!\!\!\check{D}_{\{\check{k}_n\}}\!\!\!&=&\!\!\!\sum_{n=1}^{\mrp}\check{k}_n(n-1),
\end{eqnarray}
with \eqref{multinomial_theorem_applied_SDF} obtained via the multinomial theorem \cite[Section 24.1.2]{abramowitz_stegun}. The integral $\mathcal{\check{I}}_2$ in \eqref{I_2_CogFDRS_NDL} is on the form of the Riemann-Liouville integral in \cite[Eq. 3.383-1]{gradshteyn_ryzhik} when $\check{\eta}_l <0$. Hence,
\begin{eqnarray}
\!\!\!\!\!\!\!\!\!\!\!\!\!\mathcal{\check{I}}_2
&=& \Ith^{\check{D}_{\{\check{k}_n\}}+\msp} \BTAFN{\check{D}_{\{\check{k}_n\}}+1}{\msp}
\nonumber\\&&\times
\KMRFN{\msp}{\check{D}_{\{\check{k}_n\}}+\msp+1}{-\Ith \check{\eta}_l}.
\end{eqnarray}
When $\check{\eta}_l > 0$, we can again use the change of variable $y=\Ith-\beta$ and then \cite[Eq. 3.383-1]{gradshteyn_ryzhik} to get
\begin{eqnarray}
\!\!\!\!\!\!\!\!\!\!\!\!\!\!\!\!\mathcal{\check{I}}_2
\!\!\!&=&\!\!\! e^{-\Ith \check{\eta}_l}\Ith^{\check{D}_{\{\check{k}_n\}}+\msp} \BTAFN{\msp}{\check{D}_{\{\check{k}_n\}}+1}
\nonumber\\\!\!\!\!\!\!&&\times
\KMRFN{\check{D}_{\{\check{k}_n\}}+1}{\check{D}_{\{\check{k}_n\}}+\msp+1}{\Ith \check{\eta}_l}.
\end{eqnarray}
Therefore, $\PL$ is finally given by \eqref{P_L_final}.
\begin{figure*}
\centering
\begin{eqnarray}
\!\!\!\!\!\!\!\!\!\!\PL
&=&
{K \choose L} \sum_{l=0}^L  \frac{{L \choose l}(-1)^l e^{-\frac{(K-L+l)\Ith}{\PR\thetarp}}}{\GAMFN{\msp}\left(\PS\thetasp\right)^\msp}
\sum_{\sum_{n=1}^{\mrp} \check{k}_{n} = K-L+l} \frac{\GAMFN{K-L+l+1}}{\prod_{n=1}^{\mrp}\left(\GAMFN{\check{k}_n+1}\GAMFN{n}^{\check{k}_n}\right)} \left(\frac{1}{\PR\thetarp}\right)^{\check{D}_{\{\check{k}_n\}}} \mathcal{\check{I}}_2.
\label{P_L_final}
\end{eqnarray}
\hrule
\end{figure*}

For $L=0$, $\Pzero$ is given by
\begin{eqnarray}
\!\!\!\!\!\!\!\!\!\!\!\!\!\!\!\!\!\!\Pzero
\!\!\!&=&\!\!\!\int_0^{\Ith} \left(\frac{\UPRGAMFN{\mrp,\frac{\Ith-\beta}{\PR\thetarp}}}{\GAMFN{\mrp}}\right)^{K}
\nonumber\\\!\!\!&&\!\!\!\times
\frac{\beta^{\msp-1}e^{-\frac{\beta}{\PS\thetasp}}}{\GAMFN{\msp}\left(\PS\thetasp\right)^\msp}
\mathrm{d}\beta  +\frac{\UPRGAMFN{\msp,\frac{\Ith}{\PS\thetasp}}}{\GAMFN{\msp}}.
\end{eqnarray}
Again, the integral involving higher powers of the upper regularized Gamma function is solved via its series expansion, multinomial theorem, then with the aid of the Riemann-Liouville integral in \cite[Eq. 3.383-1]{gradshteyn_ryzhik} for $\left(\frac{1}{\PS\thetasd}-\frac{K}{\PR\thetarp}\right)<0$ to yield \eqref{eq:P0_1}
\begin{figure*}
\centering
\begin{eqnarray}
\!\!\!\!\!\!\!\!\!\!\!\!\Pzero
&=& \frac{\UPRGAMFN{\msp,\frac{\Ith}{\PS\thetasp}}}{\GAMFN{\msp}} +
\frac{e^{-\frac{K\Ith}{\PR\thetarp}}}{\GAMFN{\msp}\left(\PS\thetasp\right)^\msp}
 \sum_{\sum_{n=1}^{\mrp} \acute{k}_{n} = K} \frac{\GAMFN{K+1}}{\prod_{n=1}^{\mrp}\left(\GAMFN{\acute{k}_n+1}\GAMFN{n}^{\acute{k}_n}\right)} \left(\frac{1}{\PR\thetarp}\right)^{\acute{D}_{\{\acute{k}_n\}}} \nonumber\\
&&
\times  \Ith^{\acute{D}_{\{\acute{k}_n\}}+\msp} \BTAFN{\acute{D}_{\{\acute{k}_n\}}+1}{\msp}
\times \KMRFN{\msp}{\acute{D}_{\{\acute{k}_n\}}+\msp+1}{-\Ith \left(\frac{1}{\PS\thetasd}-\frac{K}{\PR\thetarp}\right)}. \label{eq:P0_1}
\end{eqnarray}
\hrule
\end{figure*}
or, alternatively for $\left(\frac{1}{\PS\thetasd}-\frac{K}{\PR\thetarp}\right)>0$, we get \eqref{eq:P0_2}
\begin{figure*}
\centering
\begin{eqnarray}
\!\!\!\!\!\!\!\!\!\!\!\!\Pzero
&=& \frac{\UPRGAMFN{\msp,\frac{\Ith}{\PS\thetasp}}}{\GAMFN{\msp}} +
\frac{e^{-\frac{K\Ith}{\PR\thetarp}} e^{-\Ith \left(\frac{1}{\PS\thetasd}-\frac{K}{\PR\thetarp}\right)}}{\GAMFN{\msp}\left(\PS\thetasp\right)^\msp}
 \sum_{\sum_{n=1}^{\mrp} \acute{k}_{n} = K} \frac{\GAMFN{K+1}}{\prod_{n=1}^{\mrp}\left(\GAMFN{\acute{k}_n+1}\GAMFN{n}^{\acute{k}_n}\right)} \left(\frac{1}{\PR\thetarp}\right)^{\acute{D}_{\{\acute{k}_n\}}} \nonumber\\
&&
\times  \Ith^{\acute{D}_{\{\acute{k}_n\}}+\msp} \BTAFN{\msp}{\acute{D}_{\{\acute{k}_n\}}+1}
\times \KMRFN{\acute{D}_{\{\acute{k}_n\}}+1}{\acute{D}_{\{\acute{k}_n\}}+\msp+1}{\Ith \left(\frac{1}{\PS\thetasd}-\frac{K}{\PR\thetarp}\right)}. \label{eq:P0_2}
\end{eqnarray}
\hrule
\end{figure*}
with
\begin{eqnarray}
\acute{D}_{\{\acute{k}_n\}}&=&\sum_{n=1}^{\mrp}\acute{k}_n(n-1).
\end{eqnarray}

\subsection{Leveraging the Direct $\SN - \DN$ Link}
This case applies to the protocols which leverage the information that directly arrives at the destination from the source, i.e., in the \ac{IDL}/\ac{DT} and \ac{SDF} protocols. The analysis of $P_L$ follows directly from the \ac{NDL} case for $L=1,2,\cdots,K$. However, when the direct ${\SN - \DN}$ link is leveraged, communication can still take place even when no feasible relays exist. Thus, the event with probability $\Pzero$ in the \ac{NDL} case is now further split into two sub-events. Specifically, communication can still succeed in the sub-event when the sum of the source and relay transmissions do not satisfy the interference constraint for all relays but the source alone does. Let us denote the probability of this sub-event by $\tilde{\mathcal{P}}_0$. It is then given by
\begin{eqnarray}
\tilde{\mathcal{P}}_0=\mathbb{P}\left\{\left(\displaystyle\bigcap_{k=1}^K \left(\Isk>\Ith\right)\right) \cap \left(\Isp\leq \Ith\right)\right\}.
\end{eqnarray}
Note that $\Isk$ depends on $\Isp$. Nonetheless, $\{\Isk\}_{k=1}^K$ are mutually independent given $\Isp$, and distributed as in \eqref{conditional_distribution_of_interference_paths}.

Finally, when $\Isp>\Ith$, with probability $\Pzero-\tilde{\mathcal{P}}_0$, communication fails, resulting in an end-to-end \ac{SNR} of $\etoeFDSNR=0$. Hence, the distribution is a unit-step function at $\etoeFDSNR=0$. We can now write the end-to-end \ac{SNR} \ac{CDF} as
\begin{eqnarray}
\!\!\!\!\!\!\!\!\!\!F^{\rm DL}_{\etoeFDSNR}(\RVval)
\!\!\!\!\!&=&\!\!\!\!\! \Pzero-\overline{F_{\SDFDSNR}}\left(\RVval\right)\tilde{\mathcal{P}}_0
+\sum_{L=1}^K F^{\rm DL}_{\etoeFDSNR}(\RVval \vert L) \PL,
\end{eqnarray}
with ${\rm DL} \in \{\rm IDL/DT, SDF\}$. $\tilde{\mathcal{P}}_0$ is simply given as
\begin{eqnarray}
\tilde{\mathcal{P}}_0
&=& \Pzero - \frac{\UPRGAMFN{\msp,\frac{\Ith}{\PS\thetasp}}}{\GAMFN{\msp}}.
\end{eqnarray}

\section{Diversity Analysis of Full-Duplex Relay Selection Protocols}
In this section, we analyze the diversity order offered by each of the cooperative protocols under consideration.
\subsection{MHDF-NDL}
By setting $\PS=\PR=\Pgen$, and substituting the outage threshold $\SNRFD$ into the derived \ac{CDF} expressions under Rayleigh fading ($m=1$ for all fading links), the outage probability is given by
\begin{eqnarray}
\PNDL(\Pgen)
&=&\left(
1-\frac{e^{-\frac{a}{\Pgen}}}{1+b \Pgen^{\PWRSCL-1}}
\right)^K,
\end{eqnarray}
where $a=\SNRFD\left(\frac{1}{\pisr}+\frac{1}{\pird}\right)$ and $b=\SNRFD\frac{\pirr}{\pisr}$. As $\Pgen$ increases, $e^{-\frac{a}{\Pgen}}$ approaches $1-\frac{a}{\Pgen}$. Hence,
\begin{eqnarray}
\PNDL(\Pgen)
&\approx&\left(\frac{b\Pgen^\PWRSCL + a}{b \Pgen^\PWRSCL + \Pgen }\right)^K.
\end{eqnarray}
Therefore, the diversity order is given by
\begin{eqnarray}
\DNDL
\!\!\!\!&=&\!\!\!\!\lim_{\Pgen \rightarrow \infty} K \frac{\left(\frac{b + \Pgen^{1-\PWRSCL} }{b + a \Pgen^{-\PWRSCL}}\right)}{\log\left(\Pgen\right)} = K (1-\PWRSCL).
\end{eqnarray}
\subsection{MHDF-IDL}
By substituting $\PS=\PR=\Pgen$ and $\RVval=\SNRFD$ for Rayleigh fading, we get
\begin{equation}
\PIDL(\Pgen)
= \sum_{k=0}^K {K \choose k}  \frac{\left(\frac{-e^{-\frac{a}{P}}}{1+b \Pgen^{\PWRSCL -1}}\right)^k}{c_k}\approx \sum_{k=0}^K {K \choose k}  \frac{\left(\frac{a \Pgen^{-1}-1}{b \Pgen^{\PWRSCL-1}-1}\right)^k}{c_k},
\end{equation}
where $c_k=\SNRFD k\frac{\pisd}{\pird}+1$. Hence,
\begin{eqnarray}
\DIDL
=\!\!\!\!\lim_{\Pgen\rightarrow \infty} -\frac{\log\left(   \sum_{k=0}^K {K \choose k}  \frac{\left(\frac{a \Pgen^{-1}-1}{b\Pgen^{\PWRSCL-1}+1}\right)^k}{c_k}  \right)}{\log\left(\Pgen\right)}=0,\label{diversity_IDL}
\end{eqnarray}
regardless of the exact value of $0\le \PWRSCL \le 1$, since $\PIDL(\Pgen)$ is finite and not a function of $\Pgen$ as $\Pgen \rightarrow \infty$.
\subsection{MHDF-IDL/DT}
Setting $\PS=\PR=\Pgen$ and $\RVval=\SNRFD$ for Rayleigh fading yields
\begin{eqnarray}
\!\!\!\!\!\!\!\!\!\!\!\!\!\!\!\!\PMHDT(\Pgen)
&=&\sum_{k=0}^K {K \choose k} \left(\frac{-e^{-\frac{a}{\Pgen}}}{b \Pgen^{\PWRSCL-1}+1}\right)^k \frac{\left(1-e^{-\frac{d_k}{\Pgen}}\right)}{c_k}\nonumber\\
&\approx&\sum_{k=0}^K {K \choose k} \left(\frac{a-\Pgen}{b\Pgen^{\PWRSCL}+\Pgen}\right)^k \frac{\SNRFD}{\pisd}\frac{1}{\Pgen}\nonumber\\
&=&\frac{\SNRFD}{\pisd}\frac{1}{\Pgen} \left(1+\frac{a-\Pgen}{b\Pgen^{\PWRSCL}+\Pgen}\right)^K\label{P_out_hybrid_approximate}\nonumber\\
&=& \frac{\SNRFD}{\pisd}\frac{1}{\Pgen} \left(\frac{b+a\Pgen^{-\PWRSCL}}{b+\Pgen^{1-\PWRSCL}}\right)^K,
\end{eqnarray}
with $d_k=\SNRFD \left(\frac{\SNRFD k}{\pird}+\frac{1}{\pisd}\right)$, $\frac{d_k}{c_k}=\frac{\SNRFD}{\pisd}$, while \eqref{P_out_hybrid_approximate} follows from the binomial theorem. Hence, the diversity gain is given by
\begin{eqnarray}
\DMHDT
&=&\lim_{\Pgen\rightarrow \infty} -\frac{\log\left(\frac{\SNRFD}{\pisd}\frac{1}{\Pgen} \left(\frac{b+a\Pgen^{-\PWRSCL}}{b+\Pgen^{1-\PWRSCL}}\right)^K\right)}{\log\left(\Pgen\right)}\nonumber\\
&=& K(1-\PWRSCL)+1.\label{diversity_IDLDT}
\end{eqnarray}
\subsection{SDF}
Under Rayleigh fading with $\PS=\PR=\Pgen$, we have
\begin{eqnarray}
\!\!\!\!\!\!\!\!\!\!\!\!\!\!\PSDFFDRS(\Pgen)
&\approx&\sum_{k=0}^K  {K \choose k} \left(-\frac{e^{-\frac{a}{\Pgen}}}{b \Pgen^{\PWRSCL-1}+1}\right)^k \!\frac{1-e^{-\frac{p_k}{\Pgen}}}{q_k}\nonumber\\
&\approx&\sum_{k=0}^K  {K \choose k} \left(-\frac{1-\frac{a}{\Pgen}}{b \Pgen^{\PWRSCL-1}+1}\right)^k \frac{\RVval}{\PS\pisd}\frac{1}{\Pgen},\label{SDF_out_large_P}
\end{eqnarray}
where $p_k=\RVval\left(\frac{1}{\PS\pisd}-\frac{k}{\PR\pird}\right)$ and $q_k=1-k\frac{\PS\pisd}{\PR\pird}$, using the same approximation for large $\Pgen$. Clearly, this yields the same diversity order of the hybrid \ac{IDL}/\ac{DT} protocol. That is,
\begin{eqnarray}
\DSDFFDRS &=& K(1-\PWRSCL)+1.\label{diversity_SDF}
\end{eqnarray}
The derived diversity results are summarized in Table \ref{table:FDRS_diversity}.
\begin{table}[!t]
\caption{Diversity order of Full-Duplex Relay Selection Protocols with $K$ Relays}
\label{table:FDRS_diversity}
\centering
\small
\begin{tabular}{|c|c|c|}
\hline
\multicolumn{2}{ |c| }{Protocol/Scenario}			& 		Diversity Order\\
\hline
\multirow{3}{*}{\ac{MHDF}} &	\ac{NDL}		&		$K$(1~-~$\PWRSCL$)\\ \cline{2-3}
&	\ac{IDL}		&		0\\	\cline{2-3}
& 	\ac{IDL}/\ac{DT}		&		$K$(1~-~$\PWRSCL$) + 1\\
\hline
\multicolumn{2}{ |c| }{\ac{SDF}}	&$K$(1~-~$\PWRSCL$) + 1\\
\hline
\end{tabular}
\vspace{-2mm}
\end{table}

\textit{Discussion:} As clear from the diversity results in \eqref{diversity_IDL} and \eqref{diversity_IDLDT} for the \ac{IDL} and \ac{IDL}/\ac{DT}, the lost diversity gain of $K(1-\PWRSCL)$ due to treating the direct source-destination link as interference in the \ac{MHDF} protocol is interestingly recovered only by simply considering the direct link as a possible diversity branch, which also adds an additional diversity order to yield $K(1-\PWRSCL)+1$. This is due to the fact that, by considering \ac{DT}, the direct link interference realizations are now confined only to the lower range from $0$ to the outage threshold, while all higher-value realizations are utilized in direct transmission. Also, comparing \eqref{diversity_IDLDT} and \eqref{diversity_SDF}, we find that such a hybrid \ac{MHDF}-\ac{IDL}/\ac{DT} full-duplex relay selection protocol offers the same diversity order of the \ac{SDF} protocol which jointly exploits the relayed and direct paths instead of treating one as interference. Hence, although \ac{SDF} still yields better error performance, \ac{IDL}/\ac{DT} may still offer a close performance as the number of available relays increase, and hence, it may be favored for employment in relay selection scenarios at high \ac{SNR} due to its simpler decoding approach.
\section{Numerical Results and Comparisons}
We numerically evaluate the performance of opportunistic \ac{FDRS} in underlay cognitive networks, and verify the theoretical findings derived in previous sections. All numerical results are evaluated by averaging over $10^7$ sets of channel realizations with the parameters summarized in the caption of each figure. Also, for clarity of presentation, solid lines with unfilled marks are used to plot the theoretical results, while the same filled marks with no connecting lines are used for simulation results. Hence, curves with solid lines and filled marks indicate perfect matching between theoretical and simulation results.

We compare the derived and numerical results of the studied \ac{FDRS} protocols to those of \ac{HDRS} systems, exploiting the direct link whenever it exists, either using \ac{MRC} or \ac{SDF} protocols. The results for \ac{MRC}-\ac{HDRS} can be found in the literature, for instance in \cite{200912_SPL_RS_SC_Nakagami,201005_SPL_RS_Nakagami}, while that of \ac{SDF}-\ac{HDRS}, in addition to the incorporation of the primary receiver interference limit, can be easily derived using similar approaches to those in the presented manuscript. When comparing fixed-rate half-duplex and full-duplex cooperative systems, we can either compare the outage performance or the end-to-end throughput. In outage performance comparison, the rate loss in \ac{HDR} is captured via a doubled source rate. On the other hand, when throughput is compared, both full-duplex and half-duplex systems use the same source rate as that of \ac{DT}, while the \ac{HDR} rate loss is accounted for in the final throughput calculation via a multiplicative factor of half. The aforementioned approach is adopted throughout this section. Specifically, for fixed-rate transmission, the end-to-end throughput, $\TPT$, is obtained simply as
\begin{eqnarray}
\TPT = \Rate (1-\Pout) \text{ bpcu},
\end{eqnarray}
where $\Rate$ is the fixed source transmission rate in \ac{bpcu}, and $\Pout$ is the end-to-end outage probability when the source rate is equal to $\Rate$. For half-duplex cooperation, the throughput expression is simply modified to
\begin{eqnarray}
\TPT^{\rm HD} = \frac{1}{2} \Rate (1-\Pout) \text{ bpcu}.
\end{eqnarray}

\subsection{Throughput vs. Source Rate}
\begin{figure}
\centering
\begin{subfigure}[t]{\figscl\columnwidth}
\centering
\includegraphics[width=\columnwidth]{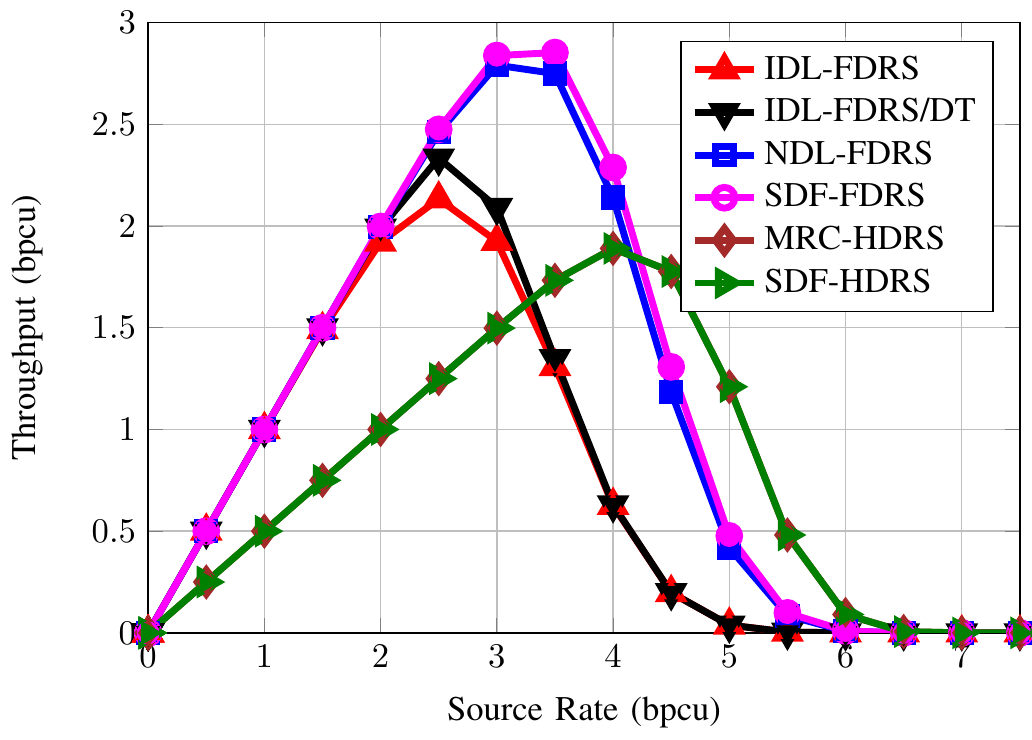}
\caption{Non-cognitive setting.}
\label{throughput_vs_rate_non_cognitive}
\end{subfigure}\\
\begin{subfigure}[t]{\figscl\columnwidth}
\includegraphics[width=\columnwidth]{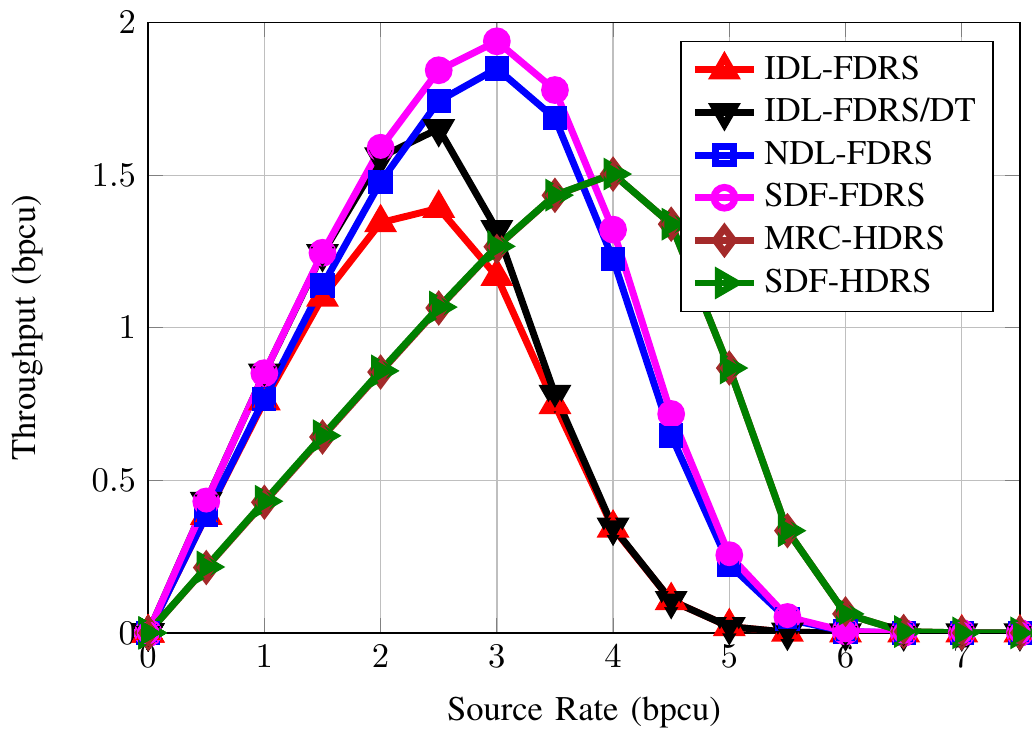}
\caption{Cognitive setting, with $\Ith=3$ dB.}
\label{throughput_vs_rate_cognitive}
\end{subfigure}
\centering
\caption{Throughput vs. rate, for ($\pisr$, $\pird$, $\pirr$, $\pisd$, $\pisp$, $\pirp$) = (15, 15, 3, 5, 0, 1) dB,  ($\msr$, $\mrd$, $\mrr$, $\msd$, $\msp$, $\mrp$) = (2, 2, 2, 2, 1, 1), $\PWRSCL=1$, $\PS=\PR=1$, and $K=3$.}
\label{throughput_vs_rate}
\vspace{-7mm}
\end{figure}
In Fig. \ref{throughput_vs_rate}, we plot the end-to-end throughput versus the source rate, for the non-cognitive and cognitive settings in Fig. \ref{throughput_vs_rate_non_cognitive} and Fig. \ref{throughput_vs_rate_cognitive}, respectively. As shown in Fig. \ref{throughput_vs_rate_non_cognitive}, \ac{FDRS} offers a remarkable throughput gain compared to \ac{HDRS}. It can be noticed that \ac{HDRS} starts to offer higher throughput only when the source rate considerably increases, which causes the adverse effect of the \ac{RSI} to increase for \ac{FDRS} while it is absent for \ac{HDRS}. It can be also noticed that the hybrid \ac{IDL}/\ac{DT} protocol uniformly outperforms that of \ac{IDL} due to exploiting the direct link, while \ac{SDF} outperforms the other \ac{FDRS} protocols whether in the presence or the absence of a direct link. In cognitive settings, as depicted in Fig. \ref{throughput_vs_rate_cognitive}, the throughput gain still exists relative to the \ac{HDRS} protocols, although the throughput is scaled down for all protocols due to the interference constraint imposed by the primary receiver. It should be also noticed that the hybrid \ac{IDL}/\ac{DT} gives a very similar performance to that of \ac{SDF} at lower values of the source rate. Next, we investigate the relative gains by varying the number of cooperating relays.

\subsection{Throughput vs. Number of Relays}
\begin{figure}
\centering
\centering
\begin{subfigure}[t]{\figscl\columnwidth}
\centering
\includegraphics[width=\columnwidth]{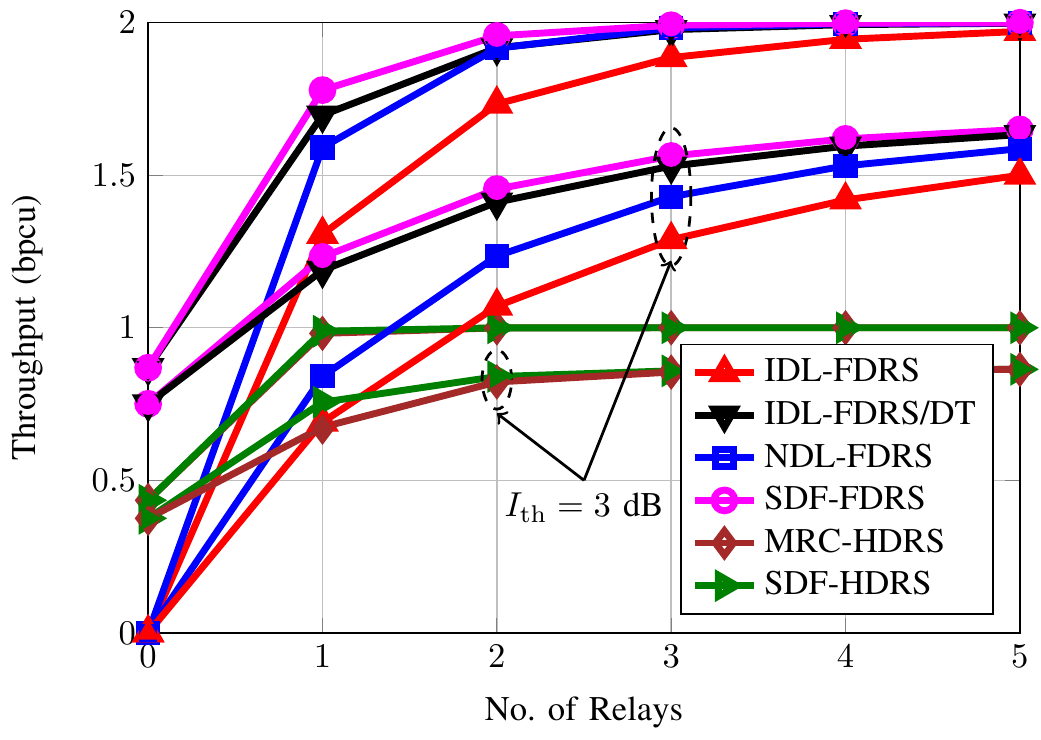}
\caption{Throughput vs. $K$.}
\label{throughput_vs_K}
\end{subfigure}\\
\begin{subfigure}[t]{\figscl\columnwidth}
\centering
\includegraphics[width=\columnwidth]{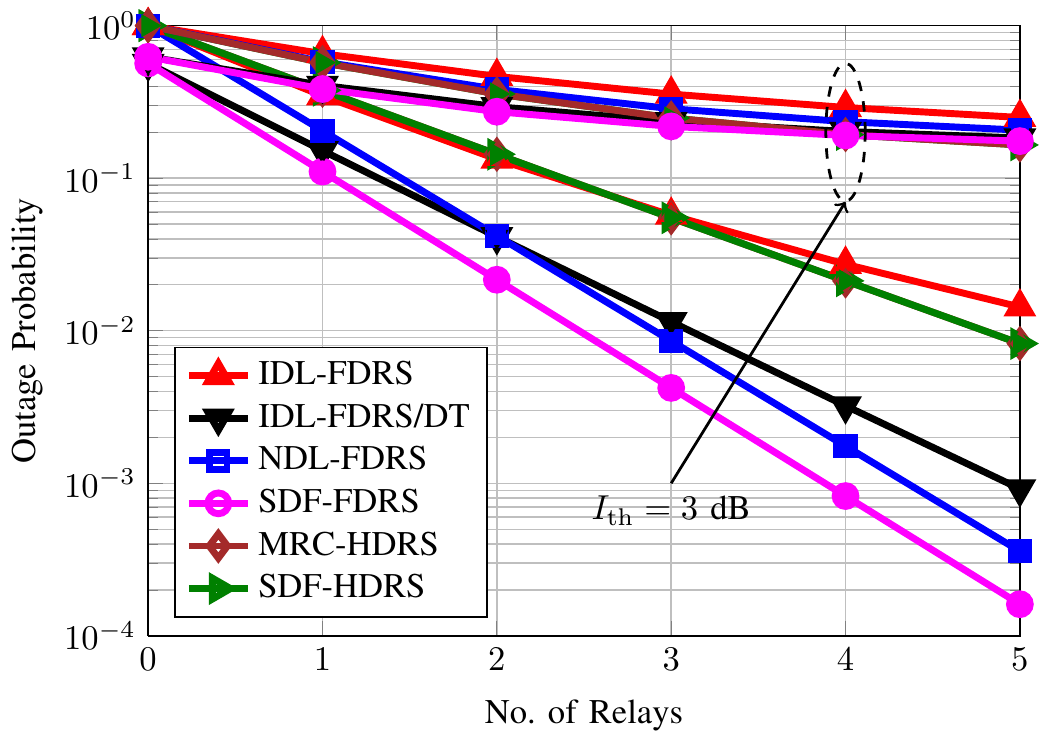}
\caption{Outage probability vs. $K$.}
\label{outage_vs_K}
\end{subfigure}
\caption{Performance vs. $K$, for ($\pisr$, $\pird$, $\pirr$, $\pisd$, $\pisp$, $\pirp$) = (15, 15, 5, 5, 0, 1) dB,  ($\msr$, $\mrd$, $\mrr$, $\msd$, $\msp$, $\mrp$) = (2, 2, 2, 2, 1, 1), $\PWRSCL=1$, $\PS=\PR=1$, and $\Rate=2$ bpcu.}
\label{performance_vs_K}
\vspace{-7mm}
\end{figure}
The end-to-end performance is shown in Fig. \ref{performance_vs_K} while varying the number of available relays. In Fig. \ref{throughput_vs_K}, the throughput gain due to the use of \ac{FDRS} is obvious for both settings where a primary receiver exists or not. The figure also emphasizes the finding that the hybrid \ac{IDL}/\ac{DT} may offer a similar performance to that of \ac{SDF}. The outage performance is also shown in Fig. \ref{outage_vs_K}.

\subsection{Outage vs. Transmit Power}
\begin{figure}
\centering
\includegraphics[width=\figscl\columnwidth]{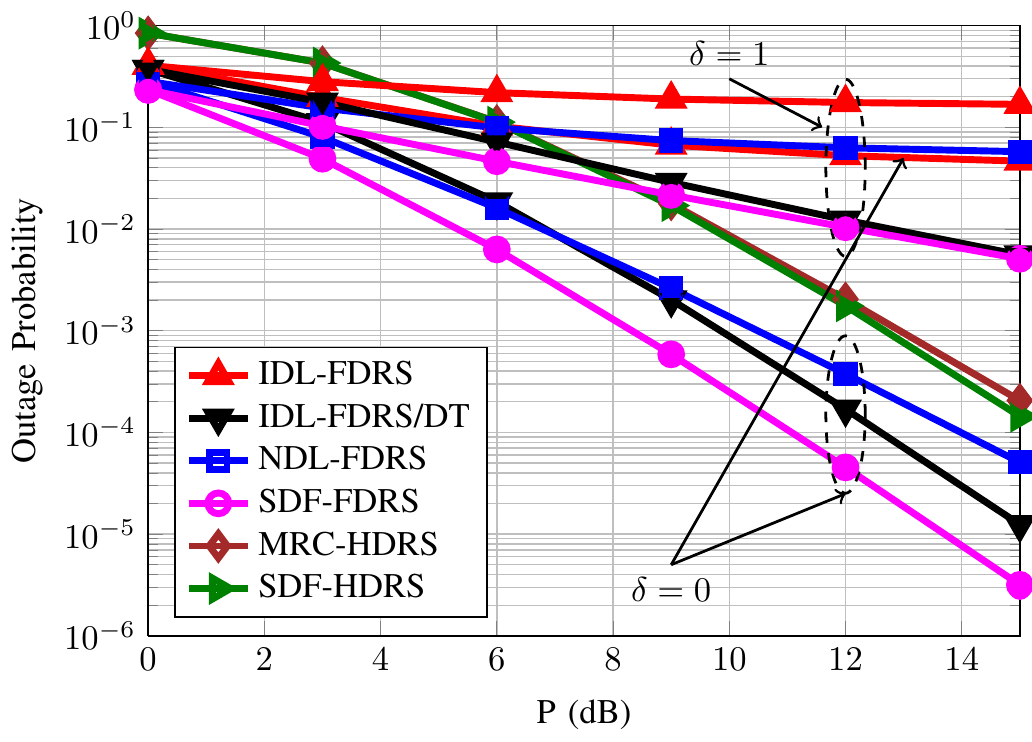}
\caption{Outage vs. $\Pgen$, for ($\pisr$, $\pird$, $\pirr$, $\pisd$) = (10, 10, 3, 0) dB,  ($\msr$, $\mrd$, $\mrr$, $\msd$) = (1, 1, 1, 1), $\PS=\PR=1$, $K=3$ relays, and $\Rate=2$ bpcu.}
\label{outage_vs_P_FDRS}
\vspace{-5mm}
\end{figure}
We show in Fig. \ref{outage_vs_P_FDRS} the outage performance of the studied protocols while varying the transmit power, $\PS=\PR=\Pgen$. This readily shows the offered diversity order of the cooperative protocols by comparing the slope of each. The diversity order of the \ac{HDRS} clearly does not depend on the value of $\PWRSCL$, with the \ac{MRC} protocol achieving a diversity of order $K=3$ while the \ac{SDF} protocol attains a diversity order of $K+1=4$. As derived in the previous section, the \ac{MHDF}-\ac{IDL}-\ac{FDRS} protocol suffers from an error floor indicating a zero diversity order regardless of the scaling trend of the \ac{RSI} link with the relay's power. The \ac{MHDF}-\ac{NDL}-\ac{FDRS} protocol performs better as it does not suffer from direct link interference, yet it still suffers from an error floor for $\PWRSCL=1$. When $\PWRSCL=0$, \ac{MHDF}-\ac{NDL}-\ac{FDRS} achieves a diversity of order $K=3$.

For the hybrid \ac{IDL}/\ac{DT} protocol, the diversity order is significantly enhanced for $\PWRSCL=0$ from $0$ to $K+1$ only by leveraging the direct link as a diversity branch on its own, even while still considering it as interference to the relayed paths. It is also enhanced from $0$ to $1$ for $\PWRSCL=1$. This result is due to the fact that the outage event caused by the direct link to all the relayed paths due to its large realizations is not only eliminated, but also successful communication can take place via \ac{DT}. Hence, only low realizations of the direct link can now contribute to the outage of the relayed paths, limiting the denominator of the \ac{SINR} for each, and recovering the diversity of each path to an order of $1$, in addition to that offered by the \ac{DT} itself. When $\PWRSCL=1$, however, each path still suffers from an error floor due to the persistence if the \ac{RSI} effect, and hence only a diversity of order $1$ is offered due to \ac{DT}. It is clear that \ac{SDF}-\ac{FDRS} offers the same diversity order for similar reasons, although it also leverages the direct link information even in the cooperative mode yielding a better error performance.
\section{Conclusion}
In this work, exact performance analysis was performed for opportunistic \ac{DF} \ac{FDRS}, taking the \ac{RSI} of the relays into account. More specifically, the exact \ac{CDF} of the end-to-end \ac{SNR} was derived considering Nakagami-$m$ fading channels. Moreover, the performance of \ac{FDRS} techniques was analyzed in underlay networks where a primary user dictates an interference threshold that cannot be exceeded by the secondary transmissions. The performance is analyzed for both relaying settings where either coverage extension or throughput enhancement is targeted. The derived exact analytical results are shown to perfectly match with those obtained via numerical simulations. Moreover, the paper's findings confirm recent results in \ac{AF} \ac{FDRS} which show that an error floor exists at high \ac{SNR} regimes in coverage extension scenarios where no direct link exists. Nonetheless, a diversity order of $1$ can be still maintained in the existence of a direct link for simple protocols that treat the direct link as interference to the relayed paths, while a diversity of order $2$ can be offered by selective \ac{FDR}. These low diversity order values apply only when the \ac{RSI} is linearly scaling with the relay's power. In the case of constant scaling, which agrees which some state-of-the-art adaptive suppression/cancellation techniques, a diversity order equal to the number of relays can be maintained.
\section*{Acknowledgment}
The first author would like to thank Prof. Ahmed K. Sultan Salem at \ac{KAUST} for the frequent fruitful discussions and the continued support. The first author would also like to thank Dr. Elsayed Ahmed at Qualcomm Inc. for the discussion on state-of-the-art results on the modeling of full-duplex radios.

\bibliographystyle{IEEEbib}
\bibliography{IEEEabrv,ref}

\end{document}